\RequirePackage{fix-cm}
\documentclass[sn-mathphys,Numbered]{sn-jnl}

\usepackage{graphicx}%
\usepackage{adjustbox}%
\usepackage{amsmath,amssymb,amsfonts}%
\usepackage{amsthm}%
\usepackage{mathrsfs}%
\usepackage[title]{appendix}%
\usepackage{xcolor}%
\usepackage{textcomp}%
\usepackage{manyfoot}%
\usepackage{booktabs}%
\usepackage{placeins}%

\hypersetup{bookmarksdepth=3}%
\renewcommand*{\theHequation}{\theHsection.\arabic{equation}}%

\theoremstyle{plain}%
\newtheorem{theorem}{Theorem}%
\newtheorem{proposition}[theorem]{Proposition}%

\theoremstyle{remark}%

\theoremstyle{definition}%

\begin{document}

\title[Hamiltonian Casimir and Static Torsion in the KV Model]%
{Gauge-Unfixed Hamiltonian Casimir and Static Torsionful Sector in the Katanaev--Volovich Model}

\author*[1]{\fnm{Jaime} \sur{Manuel Cabrera}}\email{jaime.manuel@ujat.mx}
\equalcont{These authors contributed equally to this work.}

\author[1]{\fnm{Jorge Mauricio} \sur{Paulin Fuentes}}\email{jorge.paulin@ujat.mx}
\equalcont{These authors contributed equally to this work.}

\affil*[1]{\orgdiv{Divisi\'on Acad\'emica de Ciencias B\'asicas},
\orgname{Universidad Ju\'arez Aut\'onoma de Tabasco},
\orgaddress{\street{Km 1 Carretera Cunduac\'an-Jalpa},
\city{Cunduac\'an}, \postcode{86690},
\state{Tabasco}, \country{M\'exico}}}


\abstract{We give a gauge-unfixed Hamiltonian analysis of the
first-order Katanaev--Volovich model of two-dimensional gravity with
torsion.  The first-order action already singles out natural canonical
pairs: the auxiliary Lorentz scalars are conjugate to the spatial
connection and zweibein.  An extended Dirac--Bergmann embedding isolates
an auxiliary second-class sector whose elimination recovers the canonical
structure encoded in the first-order action and leaves the first-class
constraints generating the local gauge symmetries.  An independent
Faddeev--Jackiw reduction yields the same reduced brackets.  The
Katanaev/Poisson--Sigma Casimir is then recovered, up to normalization,
directly from the reduced Dirac--Bergmann first-class constraint ideal,
before imposing any gauge condition.  This identifies the Casimir as a
global label of the reduced canonical sectors; after the static
normalization is chosen, it supplies the Hamiltonian parameter of the
torsionful branch.  The same normalization is applied to that branch in
dilaton gauge, whose field equations are verified on shell.  In this
static sector the Casimir-normalized radial field does not coincide with
the metric Killing norm: the diagonal representative satisfies \(NB=e^{4\beta r}\).
Consequently, torsion modifies the Killing temperature through the
normalization of the Killing time, while the horizon entropy retains its
standard two-dimensional dilaton value and the Casimir-normalized first
law holds.}

\keywords{Two-dimensional gravity; Katanaev--Volovich model;
gravity with torsion; Hamiltonian reduction; Poisson--Sigma models;
black-hole thermodynamics}

\maketitle


\section{Introduction}\label{sec1}

Two-dimensional gravitational theories provide a controlled setting in
which the interplay between gauge symmetry, integrability, constraints
and global observables can be analyzed explicitly.  Although the pure
Einstein--Hilbert action in two dimensions is topological and gives no
local field equations, nontrivial models arise once one enlarges the
geometric structure or introduces additional fields.  A central example is the Katanaev--Volovich (KV) model: it is formulated
in Riemann--Cartan geometry and contains quadratic terms in both curvature
and torsion, together with a cosmological constant
\cite{KatanaevVolovich1986PLB,katanaev1990complete}.

The KV model has a well-established place in the literature on
two-dimensional gravity.  The nonlinear gauge-theory formulation of
related two-dimensional dilaton-gravity models was developed by Ikeda and
Izawa, while the terminology and geometric language of Poisson--Sigma
models were subsequently adopted and systematized by Schaller and Strobl
\cite{ikeda1993quantum,Ikeda1994_NonlinearGauge2D,SchallerStrobl1994}.
The classical integrability, solution space and canonical formulations of
the KV model have been studied in different gauges and variables
\cite{Katanaev1988_TMP_Hamiltonian,katanaev1990complete,strobl1993all,%
grignani1993canonical,kummer1993comment,katanaev2002effective}.  In
particular, the local integrability and explicit solution structure are
part of the established literature.  The present work does not propose a
new local solution of the field equations; it refines the canonical
description of the established solution space in two complementary
directions.  First, it gives a
\emph{gauge-unfixed} Hamiltonian reconstruction of the KV integrable
structure.  The extended Dirac--Bergmann phase space isolates the
auxiliary second-class sector generated by the embedding, and its
elimination yields the reduced Dirac brackets.  These brackets are then
compared with the Faddeev--Jackiw symplectic brackets.  The
Katanaev/Poisson--Sigma Casimir is recovered separately from the reduced
Dirac--Bergmann first-class constraint ideal.  In this reconstruction,
the Casimir labels global sectors of the reduced theory, up to
conventional normalization, rather than appearing as an additional
constraint in the Dirac--Bergmann consistency chain.  Second, the same
canonical normalization is applied to a static torsionful sector.  For
\(\beta\neq0\), the Casimir-normalized field, denoted by \(\xi\)
in the static analysis, fixes the radial metric function through
\(B=\xi^{-1}\), whereas the metric Killing norm determines the surface
gravity.  This separation gives \(N=e^{4\beta r}\xi\) and
\(NB=e^{4\beta r}\), a torsion-dependent Killing temperature and a
Casimir-normalized first law with the standard two-dimensional dilaton
entropy.  The construction links the local constrained dynamics with the
global Casimir label and with the thermodynamic normalization of the
static sector in two-dimensional gravity
\cite{GrumillerKummer2000,GrumillerKummerVassilevich2002}.

The extended phase space used below is a methodological device, not a
claim that the first-order action lacks canonical pairs.  After a $1+1$
decomposition its kinetic term is
\(\varphi\,\dot\omega_x+\varphi_I\,\dot e^I_x\), so that \(\varphi\) and
\(\varphi_I\) are the momenta conjugate to \(\omega_x\) and \(e^I_x\)
(developed in Section~\ref{sec:VK}).  Treating \(\varphi,\varphi_I\)
as configuration variables introduces the auxiliary second-class sector,
which implements these canonical identifications through Dirac brackets.
This embedding makes the Dirac--Bergmann reduction comparable, in the same
variables, with the modified Faddeev--Jackiw reduction and with the
gauge-unfixed construction of the Casimir from the reduced first-class
constraint ideal.

The approach adopted here is also distinct from Hamiltonian treatments
performed after imposing the conformal gauge.  In Katanaev's
conformal-gauge analysis, the residual constraints are imposed as
subsidiary conditions inherited from the original covariant equations
\cite{Katanaev1988_TMP_Hamiltonian}.  By contrast, the present analysis
is carried out before gauge fixing: the primary constraints are obtained
from the singular Legendre map of the unreduced first-order action, and
the remaining constraints follow from their time preservation.

The paper is organized as follows.  Section~\ref{sec:VK} fixes the
conventions, introduces the first- and second-order forms of the KV
action, and explains the distinction between the natural canonical
reading and the extended phase-space embedding.  Section~\ref{sec:Dirac}
develops the Dirac--Bergmann constraint analysis.  Section~\ref{sec:first-second-class}
separates the first- and second-class sectors, constructs the reduced
Dirac brackets, and closes the local gauge analysis through the extended
Hamiltonian and gauge generator.  Section~\ref{sec:FJ} implements the
Faddeev--Jackiw reduction and verifies the bracket equivalence.
Section~\ref{sec:casimir} derives the gauge-unfixed Casimir from the
reduced constraint ideal.  Section~\ref{sec:BH} applies this Casimir to a
static torsionful sector and derives the corresponding horizon,
temperature and first-law relations.  Section~\ref{sec:disc} summarizes
the results and outlines controlled extensions involving boundary
conditions and global charges.

\section{The Katanaev--Volovich model of two-dimensional gravity}\label{sec:VK}

We briefly review the two-dimensional gravity model with dynamical
torsion introduced by Katanaev and Volovich within the Riemann--Cartan
framework \cite{KatanaevVolovich1986PLB}.

\subsection{Conventions and notation}\label{subsec:conventions}

We use spacetime coordinates $x^\mu=(t,x)$ and internal Lorentz indices
$I,J,\ldots=0,1$.  Internal indices are raised and lowered with
$\eta_{IJ}=\mathrm{diag}(+,-)$, whereas spacetime indices are moved with
$g_{\mu\nu}=\eta_{IJ}e^I{}_{\mu}e^J{}_{\nu}$.  We follow the
orientation convention used in the Katanaev--Volovich/Ikeda--Izawa and
Kummer--Strobl literature: the internal antisymmetric symbol is
normalized by
\begin{equation}
\varepsilon_{01}=+1,\qquad \varepsilon^{01}=-1,
\label{epsilon-convention}
\end{equation}
so that $\varepsilon^{10}=+1$.  Mixed components are obtained by
$\eta$-raising and lowering,
\begin{equation}
\varepsilon^I{}_{J}=\eta^{IK}\varepsilon_{KJ},
\qquad
\varepsilon_I{}^{J}=\varepsilon_{IK}\eta^{KJ}.
\label{mixed-epsilon-convention}
\end{equation}
The spacetime density is denoted by $\epsilon^{\mu\nu}$, with
$\epsilon^{tx}=+1$; equivalently, $\epsilon^{tr}=+1$ in the dilaton
gauge used later.  Thus $\epsilon$ is reserved for spacetime densities,
whereas $\varepsilon$ denotes internal Lorentz antisymmetric objects.
With these choices,
\begin{equation}
e=\det e^I{}_{\mu}
=\frac12\epsilon^{\mu\nu}\varepsilon_{IJ}e^I{}_{\mu}e^J{}_{\nu}.
\label{zweibein-determinant-convention}
\end{equation}
This convention fixes the signs in the torsion, the secondary
constraints, the Casimir representative and the static-sector
thermodynamic quantities used below.

\subsection{Second-order and first-order forms of the action}\label{subsec:action-forms}

The theory is formulated in terms of a zweibein and Lorentz connection
$(e^{I}_{\mu},\omega_{\mu})$, with action
\begin{align}
S[e^{I}_{\mu},\omega_{\mu}]=\int d^{2}x\, e\biggl(
\frac{1}{16\alpha}R{^{IJ}}_{\mu\nu}R{^{\mu\nu}}_{IJ}
-\frac{1}{8\beta}T{^{I}}_{\mu\nu}T{^{\mu\nu}}_{I}-\Lambda \biggr),
\label{1aa}
\end{align}
where $\alpha, \beta, \Lambda$ are constants, and $R{^{IJ}}_{\mu\nu}$
and $T{^{I}}_{\mu\nu}$ denote the curvature and torsion tensors,
respectively.  With the conventions fixed in Section~\ref{subsec:conventions}, we use
\begin{align}
e &= \det e^{I}_{\mu}, \nonumber \\
R{^{IJ}}_{\mu\nu} &= \varepsilon^{IJ}R_{\mu\nu},\quad
R_{\mu\nu}=\partial_{\mu}\omega_{\nu}-\partial_{\nu}\omega_{\mu},
\nonumber\\
T{^{I}}_{\mu\nu} &= \partial_{\mu}e^{I}_{\nu}
+\varepsilon^{IJ}\omega_{\mu}e_{\nu J}-(\mu\leftrightarrow\nu).
\label{1aaa}
\end{align}
The action~\eqref{1aa} follows the first-order/Yang--Mills-like
normalization used in the Ikeda--Izawa formulation, with
$\Lambda$ denoting the cosmological parameter in our notation
\cite{ikeda1993quantum,Ikeda1994_NonlinearGauge2D}.  It is equivalent,
up to conventional reparametrizations of couplings and orientation
symbols, to the Katanaev--Volovich quadratic curvature--torsion model
\cite{KatanaevVolovich1986PLB}.

The action~\eqref{1aa} belongs to the standard class of
reparametrization-invariant two-dimensional functionals quadratic in
curvature and torsion that yield second-order field equations for the
zweibein and Lorentz connection \cite{KatanaevVolovich1986PLB}.

The integrability of the model and its solution space have been analyzed
in several formulations.  In the conformal gauge, Katanaev showed that
the equations reduce to a constrained Hamiltonian system with residual
first-class constraints \cite{Katanaev1988_TMP_Hamiltonian}.  Later work
established explicit general solutions for broader two-dimensional
curvature--torsion systems, including the quadratic KV case, and clarified
the role of the corresponding conserved quantities
\cite{katanaev1990complete,katanaev2002effective}.

For $e\neq0$, the action~\eqref{1aa} can be written as
\begin{align}
S[e^{I}_{\mu},\omega_{\mu}]=\int d^{2}x\biggl(
\frac{1}{4e\alpha}F^{2}+\frac{1}{4e\beta}T_{I}T^{I}-e\Lambda\biggr),
\label{1b}
\end{align}
with
\begin{align}
F=\frac{1}{2}\epsilon^{\mu\nu}R_{\mu\nu}, \qquad
T^{I}=\frac{1}{2}\epsilon^{\mu\nu}T^{I}_{\mu\nu},
\label{1bb}
\end{align}
and it admits the first-order representation
\begin{align}
\tilde{S}[e^{I}_{\mu},\omega_{\mu},\varphi,\varphi_{I}]
=\int d^{2}x \left(\varphi F + \varphi_{I}T^{I}
-e(\alpha\varphi^{2}+\beta\varphi_{I}\varphi^{I}+\Lambda)\right).
\label{1c}
\end{align}
The equations of motion for $(e^{I}_{\mu},\omega_{\mu},\varphi,\varphi_{I})$
are
\begin{align}
\frac{\delta\tilde{S}}{\delta e^{I}_{\mu}}&:&
\epsilon^{\mu\nu}\!\left(\partial_{\nu}\varphi_{I}
+\varepsilon_{IJ}\omega_{\nu}\varphi^{J}
-\varepsilon_{IJ}e^{J}_{\nu}
(\alpha\varphi^{2}+\beta\varphi_{K}\varphi^{K}+\Lambda)\right)=0,
\nonumber\\
\frac{\delta\tilde{S}}{\delta\omega_{\mu}}&:&
\epsilon^{\mu\nu}\!\left(\partial_{\nu}\varphi
+\varepsilon^{IJ}\varphi_{I}e_{\nu J}\right)=0,\nonumber \\
\frac{\delta\tilde{S}}{\delta\varphi}&:&
F-2\alpha e\varphi=0,\nonumber\\
\frac{\delta\tilde{S}}{\delta\varphi_{I}}&:&
T^{I}-2\beta e\varphi^{I}=0.
\label{1bbb}
\end{align}
Eliminating $\varphi$ and $\varphi_{I}$ from~\eqref{1c} by means of
their equations of motion reproduces~\eqref{1b}.  For $\beta=0$, the
fields $\varphi_{I}$ act as Lagrange multipliers imposing $T^{I}=0$, and
one is left with a torsionless $R^{2}$ theory with cosmological constant.
More generally, for suitable choices of $\alpha$, $\beta$ and $\Lambda$
one recovers other familiar models of two-dimensional pure gravity
\cite{ikeda1993quantum,katanaev1994canonical}.

The formulae obtained below at $\beta=0$ are understood at the level of
the first-order action~\eqref{1c}.  In that formulation, setting
$\beta=0$ turns $\varphi_I$ into a Lagrange multiplier imposing $T^I=0$,
and the model reduces to a torsionless $R^2$-type dilaton-gravity sector.
This limit is distinct from a direct substitution $\beta=0$ in the
second-order action~\eqref{1aa}, where the torsion-squared term carries
the coefficient $1/\beta$.  In Section~\ref{sec:BH}, the $\beta=0$
expressions serve as consistency checks of the first-order formulation
and of the Casimir normalization.

\subsection{Natural canonical reading and extended phase-space
embedding}\label{subsec:naturalcanonical}

The first-order representation~\eqref{1c} is already adapted to
Hamiltonian analysis.  After the $1+1$ split performed below, its
symplectic potential contains
\begin{equation}
\Theta=\int_{\Sigma}dx\,\left(\varphi\,\delta\omega_x
+\varphi_I\,\delta e^I_x\right),
\label{canonical-potential}
\end{equation}
which suggests the minimal canonical identifications
\begin{equation}
\pi_{\omega_x}=\varphi,\qquad \pi^x_I=\varphi_I.
\label{minimal-canonical-pairs}
\end{equation}
In this reduced canonical reading, $\omega_t$ and $e^I_t$ play the role
of Lagrange multipliers enforcing the Gauss-type constraints of the
model.

The analysis below uses a larger phase space in which $\varphi$ and
$\varphi_I$ are also treated as configuration variables.  This produces
the primary constraints
\begin{equation}
\Pi\approx0,\qquad
\Pi_I\approx0,\qquad
\Pi^x-\varphi\approx0,\qquad
\Pi^x_I-\varphi_I\approx0,
\label{auxiliary-second-class-preview}
\end{equation}
which form an auxiliary second-class sector.  They implement the
canonical identifications~\eqref{minimal-canonical-pairs} through Dirac
brackets.  Once this sector is eliminated, the bracket structure reduces
to the natural canonical structure encoded in~\eqref{canonical-potential}.
Thus the extended embedding is not introduced because the first-order
action lacks canonical pairs.  It is introduced to display, within the
standard Dirac--Bergmann algorithm, how the auxiliary second-class sector
enforces the canonical identifications already present in the first-order
symplectic potential and how the resulting reduced structure compares
with the Faddeev--Jackiw reduction in the same variables.

\section{Hamiltonian constraint analysis}\label{sec:Dirac}

To develop the Hamiltonian analysis, we assume a spacetime topology
$\mathbb{R}\times\Sigma$, where $\mathbb{R}$ represents the time
evolution parameter and $\Sigma$ is a one-dimensional spatial manifold
of arbitrary but fixed topology.  The topology is held fixed throughout
the local Hamiltonian analysis; boundary conditions, nontrivial global
sectors and finite charge questions require additional data and are
therefore left to the extensions discussed in Section~\ref{sec:disc}.
Performing the $1+1$ decomposition,
the action~\eqref{1c} becomes
\begin{align}
\tilde{S}=\int d^{2}x\,\bigl(\varphi\dot{\omega}_{x}
+\varphi_{I}\dot{e}^{I}_{x}
+(\partial_{x}\varphi+\varepsilon{^{I}}_{J}e^{J}_{x}\varphi_{I})\omega_{t}
+(D_{x}\varphi_{I})e^{I}_{t}-eE\bigr),
\label{15a}
\end{align}
where $E=\alpha\varphi^{2}+\beta\varphi_{I}\varphi^{I}+\Lambda$ and
$D_{x}\varphi_{I}=\partial_{x}\varphi_{I}+\varepsilon{_{I}}^{J}
\omega_{x}\varphi_{J}$.

The minimal canonical pairs~\eqref{minimal-canonical-pairs} could be used
from the outset.  We keep instead the extended variables
$\varphi,\varphi_I$ and their conjugate momenta, so that the auxiliary
second-class sector and its elimination can be compared explicitly with
the Faddeev--Jackiw symplectic reduction.

The momenta $\Pi^{Q_{L}}=(\Pi,\Pi_{I},\Pi^{\mu}_{I},\Pi^{\mu})$
conjugate to $Q_{L}=(\varphi,\varphi^{I},e^{I}_{\mu},\omega_{\mu})$
are determined by
\begin{equation}
\Pi^{Q_{L}}=\frac{\delta{\mathcal{L}}}{\delta\dot{Q_{L}}}.
\label{16}
\end{equation}
Since the Hessian matrix $H_{LM}=\partial^2\mathcal{L}/
(\partial\dot{Q}_{L}\partial\dot{Q}_{M})=0$ has rank zero, the Legendre
map yields nine primary constraints.  From the momentum definitions~\eqref{16}:
\begin{equation}\label{18}
\begin{aligned}
\Phi &:= \Pi \approx 0, &\qquad
\Phi_{I} &:= \Pi_{I} \approx 0, &\qquad
\Phi^{t}_{I} &:= \Pi^{t}_{I} \approx 0, \\
\Phi^{x}_{I} &:= \Pi^{x}_{I}-\varphi_{I}\approx0, &\qquad
\Phi^{t} &:= \Pi^{t}\approx0, &\qquad
\Phi^{x} &:= \Pi^{x}-\varphi\approx0.
\end{aligned}
\end{equation}
The symbol $\approx$ denotes weak equality in the sense of Dirac.  The
subset $(\Phi,\Phi_I,\Phi^x,\Phi^x_I)$ is auxiliary in the sense
explained in Section~\ref{subsec:naturalcanonical}: after its
elimination, $\varphi$ and $\varphi_I$ become the momenta conjugate to
$\omega_x$ and $e^I_x$.

Performing a Legendre transformation yields the canonical Hamiltonian
\begin{equation}
H_c=-\int dx\,\bigl((\partial_{x}\varphi
+\varepsilon{^{I}}_{J}e^{J}_{x}\varphi_{I})\omega_{t}
+(\partial_{x}\varphi_{I}+\varepsilon{_{I}}^{J}\omega_{x}\varphi_{J}
-\varepsilon_{IJ}e^{J}_{x}E)e^{I}_{t}\bigr),
\label{19}
\end{equation}
and the primary Hamiltonian is
\begin{equation}
H_P=H_c+\int dx\left[
\lambda\Phi+\lambda^{I}\Phi_{I}+\lambda^{I}_{\mu}\Phi^{\mu}_{I}
+\lambda_{\mu}\Phi^{\mu}\right],
\label{20}
\end{equation}
where $(\lambda,\lambda^{I},\lambda^{I}_{\mu},\lambda_{\mu})$ are
Lagrange multipliers.  The fundamental Poisson brackets are
\begin{align}
\{\varphi(x),\Pi(y)\} &= \delta(x-y), \nonumber \\
\{\varphi^{I}(x),\Pi_{J}(y)\} &= \delta^{I}_{J}\delta(x-y), \nonumber \\
\{e^{I}_{\mu}(x),\Pi^{\nu}_{J}(y)\} &=
\delta^{I}_{J}\delta^{\nu}_{\mu}\delta(x-y), \nonumber \\
\{\omega_{\mu}(x),\Pi^{\nu}(y)\} &= \delta^{\nu}_{\mu}\delta(x-y).
\label{21}
\end{align}

The non-vanishing Poisson brackets among the constraints~\eqref{18} are
\begin{align}
\{\Phi^{x}(x),\Phi(y)\} &= -\delta(x-y), \nonumber \\
\{\Phi^{x}_{I}(x),\Phi_{J}(y)\} &= -\eta_{IJ}\delta(x-y),
\label{22}
\end{align}
which in matrix form read
\begin{align}
W=\{\widetilde{\Phi}^{A}(x),\widetilde{\Phi}^{B}(y)\}=
\left(
\begin{array}{cccccc}
  0&0&0&0&0&1\\
  0&0&0&\eta_{IJ}&0&0\\
  0&0&0&0&0&0\\
  0&-\eta_{IJ}&0&0&0&0\\
  0&0&0&0&0&0\\
  -1&0&0&0&0&0\\
\end{array}
\right)\delta(x-y),
\label{22a}
\end{align}
where $\widetilde{\Phi}^{A}=(\Phi,\Phi_{I},\Phi^{\mu}_{I},\Phi^{\mu})$.
The matrix has rank six and admits three linearly independent null
vectors.  Contracting these null vectors and enforcing the time
preservation of $\Phi^{t}_{I}$ and $\Phi^{t}$ yields the three
secondary constraints
\begin{align}
\dot{\Phi}^{t}_{I} &= \{\Phi^{t}_{I}(x),H_{P}\}\approx0
\quad\Rightarrow\quad
\Psi_{I}:=\partial_{x}\varphi_{I}
+\varepsilon{_{I}}^{J}\omega_{x}\varphi_{J}
-\varepsilon_{IJ}e^{J}_{x}E\approx0, \nonumber \\
\dot{\Phi}^{t} &= \{\Phi^{t}(x),H_{P}\}\approx0
\quad\Rightarrow\quad
\Psi:=\partial_{x}\varphi
+\varepsilon{^{I}}_{J}e^{J}_{x}\varphi_{I}\approx0,
\label{23}
\end{align}
and the rank allows us to fix the following values for the Lagrange
multipliers:
\begin{align}
\dot{\Phi}&=\{\Phi(x),H_{P}\}\approx0 \Rightarrow
\lambda_{x}-\partial_{x}\omega_{t}
-2\alpha\varphi\varepsilon_{IJ}e^{I}_{t}e^{J}_{x}\approx0, \nonumber \\
\dot{\Phi}_{I}&=\{\Phi_{I}(x),H_{P}\}\approx0 \Rightarrow
\lambda^{I}_{x}-\partial_{x}e^{I}_{t}
-\varepsilon{^{I}}_{J}\omega_{x}e^{J}_{t}
+\varepsilon{^{I}}_{J}e^{J}_{x}\omega_{t}
-2\beta\varepsilon_{JK}\varphi^{I}e^{J}_{t}e^{K}_{x}\approx0,
\nonumber \\
\lambda &= \text{undetermined}, \nonumber \\
\lambda^{I} &= \text{undetermined}.
\label{24}
\end{align}
Consistency requires the secondary constraints to be conserved in time;
for this theory no tertiary constraints arise.  The constraints in
\eqref{23} are consistent with those obtained in
\cite{akdeniz1992canonical,strobl1993all} for related forms of the KV
action; the key distinction is that the present analysis is performed
on the first-order action~\eqref{1c} without prior gauge fixing.

\section{Reduced Dirac structure and gauge symmetries}\label{sec:first-second-class}

\subsection{First- and second-class constraints}

The Dirac consistency algorithm has closed with the secondary constraints
\eqref{23} and with no tertiary constraints.  We now classify the complete
constraint set into first- and second-class sectors.  This classification
is performed before imposing any gauge condition and is based on the
Poisson-bracket matrix of all primary and secondary constraints.

With the explicit ordering
\[
(\Phi,\Phi_J,\Phi^t_J,\Phi^x_J,\Phi^t,\Phi^x,\Psi,\Psi_J),
\]
we define
\begin{align}
W_E=\{\widetilde{\Phi}^{*A}(x),\widetilde{\Phi}^{*B}(y)\},
\label{24a}
\end{align}
where \(\widetilde{\Phi}^{*A}\) denotes the corresponding
twelve-component constraint vector.  The non-vanishing Poisson brackets
are
\begin{align}
\{\Phi^{x}(x),\Phi(y)\}&= -\delta(x-y), \nonumber \\
\{\Phi^{x}_{I}(x),\Phi_{J}(y)\}&= -\eta_{IJ}\delta(x-y), \nonumber\\
\{\Phi(x),\Psi_{J}(y)\}&=
2\alpha\varphi\varepsilon_{JK}e^{K}_{y}\delta(x-y), \nonumber\\
\{\Phi(x),\Psi(y)\}&= -\partial_{y}\delta(x-y),\nonumber\\
\{\Phi_{I}(x),\Psi_{J}(y)\}&=
-\partial_{y}\delta(x-y)+\varepsilon_{IJ}\omega_{y}\delta(x-y)
+2\beta\varepsilon_{JK}\varphi_{I}e^{K}_{y}\delta(x-y),\nonumber\\
\{\Phi_{I}(x),\Psi(y)\}&=
-\varepsilon_{IJ}e^{J}_{y}\delta(x-y), \nonumber \\
\{\Phi^{x}_{I}(x),\Psi_{J}(y)\}&=
-\varepsilon_{IJ}E\delta(x-y), \nonumber \\
\{\Phi^{x}_{I}(x),\Psi(y)\}&=
\varepsilon{_{I}}^{J}\varphi_{J}\delta(x-y), \nonumber \\
\{\Phi^{x}(x),\Psi_{J}(y)\}&=
-\varepsilon{_{J}}^{K}\varphi_{K}\delta(x-y).
\label{25}
\end{align}

These brackets determine the rank and nullity of the complete constraint
matrix.  The computation is most transparent in smeared form, with
derivatives of the delta distribution integrated by parts onto the
corresponding test functions.  In this form, the auxiliary block associated
with \((\Phi,\Phi_I,\Phi^x,\Phi^x_I)\) is nonsingular and has rank six.
The remaining six independent directions are null directions of the full
matrix once the secondary constraints are combined with the auxiliary
sector.  Equivalently,
\begin{equation}
\operatorname{rank}W_E=6,\qquad
\operatorname{nullity}W_E=6 .
\label{rank-nullity-WE}
\end{equation}
Appendix~\ref{app:rank-check} records the corresponding smeared rank
check.

The six null directions are represented by four tensorial families: a
Lorentz-scalar direction \(V^{(1)}\), an internal-vector family
\(V^{(2)}{}_{I}\), a temporal internal-vector family \(V^{(3)}{}_{I}\),
and a temporal scalar direction \(V^{(4)}\).  The distinction is
componential.  The object \(V^{(1)}\) has no free internal index, although
some of its entries carry an index \(J\) because they sit in the
\(\Phi_J\) and \(\Phi^x_J\) slots of the constraint vector.  By contrast,
\(V^{(2)}{}_{I}\) and \(V^{(3)}{}_{I}\) carry a genuine free internal
index \(I=0,1\), and therefore each represents two independent null
directions.  The total count is thus \(1+2+2+1=6\).  A representative set
of null vectors is\footnote{Here \(\delta\) stands for \(\delta(x-y)\) in this display.}
\begin{equation}
\begin{aligned}
V^{(1)}={}&\bigl(0,\varepsilon^{JL}\varphi_{L}\delta,0,
\varepsilon^{J}{}_{K}e^{K}_{x}\delta,0,
-\partial_{x}\delta,0,\delta\bigr), \\
V^{(2)}{}_{I}={}&\bigl(-\varepsilon_{I}{}^{K}\varphi_{K}\delta,
-\varepsilon^{J}{}_{I}E\delta,0,
-\delta^{J}_{I}\partial_{x}\delta
-\varepsilon^{J}{}_{I}\omega_{x}\delta \\
&\quad{}-2\beta\varepsilon_{IK}\varphi^{J}e^{K}_{x}\delta,
0,-2\alpha\varepsilon_{IJ}\varphi e^{J}_{x}\delta,
\delta^{J}_{I}\delta,0\bigr), \\
V^{(3)}{}_{I}={}&\bigl(0,0,\delta^{J}_{I}\delta,0,0,0,0,0\bigr), \\
V^{(4)}={}&\bigl(0,0,0,0,\delta,0,0,0\bigr).
\end{aligned}
\label{null-vectors-display}
\end{equation}

Here the indices \(J,K,L\) appearing in the entries of \(V^{(1)}\) are
component labels associated with the vectorial constraint slots and do not
label distinct null vectors.  Thus \(V^{(1)}\) gives a single
Lorentz-scalar direction.  The family \(V^{(2)}{}_{I}\) gives the two
components of the internal first-class combination \(\gamma_I\).  The
family \(V^{(3)}{}_{I}\) has support only on the temporal primary
constraints \(\Phi^t_I\) and gives the two temporal internal-vector
directions \(\gamma^t_I=\Phi^t_I\).  Finally, \(V^{(4)}\) has support only
on the temporal primary constraint \(\Phi^t\), giving the temporal scalar
direction \(\gamma^t=\Phi^t\).

Contracting these null directions with the full constraint vector yields
the six first-class combinations
\[
\gamma,\qquad \gamma_I,\qquad \gamma^t_I,\qquad \gamma^t ,
\]
with component count \(1+2+2+1=6\).  This is the same first-class set
reported in Table~\ref{tab:constraint-classification}; the remaining
constraints form the second-class auxiliary sector
\((\chi,\chi_I,\chi^x,\chi^x_I)\).

Explicitly, the six first-class constraints are
\begin{align}
\gamma &= \Psi+\partial_{x}\chi^{x}
+\varepsilon^{JL}\varphi_{L}\chi_{J}
+\varepsilon{^{J}}_{K}e^{K}_{x}\chi^{x}_{J}\approx0, \nonumber \\
\gamma_{I}&=
\Psi_{I}+D_{x}\chi^{x}_{I}
-2\beta\varepsilon_{IK}e^{K}_{x}\varphi^{J}\chi^{x}_{J}
-2\alpha\varphi\varepsilon_{IK}e^{K}_{x}\chi^{x}
-\varepsilon{_{I}}^{K}\varphi_{K}\chi
+\varepsilon{_{I}}^{J}E\chi_{J}\approx0, \nonumber\\
\gamma^{t}_{I} &= \Phi^{t}_{I}=\Pi^{t}_{I}\approx0, \nonumber\\
\gamma^{t} &= \Phi^{t}=\Pi^{t}\approx0,
\label{26}
\end{align}
and the six second-class constraints are
\begin{equation}\label{27}
\begin{aligned}
\chi &:= \Pi\approx0, &\qquad
\chi_{I} &:= \Pi_{I}\approx0, \\
\chi^{x}_{I} &:= \Pi^{x}_{I}-\varphi_{I}\approx0, &\qquad
\chi^{x} &:= \Pi^{x}-\varphi\approx0.
\end{aligned}
\end{equation}
The second-class constraints~\eqref{27} enforce the canonical
identifications~\eqref{minimal-canonical-pairs} through Dirac brackets.
They do not impose additional dynamical restrictions.  The first-class
combinations~\eqref{26} encode the local gauge content of the model.
Table~\ref{tab:constraint-classification} summarizes the component
count used in the local degree-of-freedom calculation.

\begin{table}[!htbp]
\caption{Constraint classification in the extended Dirac--Bergmann phase
space.  Internal indices $I=0,1$ are counted explicitly.  The auxiliary
second-class sector only enforces the natural canonical identifications
of the first-order action.}
\label{tab:constraint-classification}
\centering
\begin{tabular}{lllp{0.42\textwidth}}
\toprule
Sector & Constraints & Components & Role \\
\midrule
First class & $\gamma$ & 1 & Lorentz-type local gauge generator \\
First class & $\gamma_I$ & 2 & Translation/diffeomorphism-related generators \\
First class & $\gamma^t$ & 1 & Primary multiplier constraint for $\omega_t$ \\
First class & $\gamma^t_I$ & 2 & Primary multiplier constraints for $e^I_t$ \\
Second class & $\chi,\chi^x$ & 2 & Enforce $\Pi^x=\varphi$ \\
Second class & $\chi_I,\chi^x_I$ & 4 & Enforce $\Pi^x_I=\varphi_I$ \\
\midrule
Total & & $6+6$ & $N_{\rm dof}=\tfrac12(18-2\times6-6)=0$ \\
\bottomrule
\end{tabular}
\end{table}
\FloatBarrier

The table gives the component count entering the local degree-of-freedom
calculation.  To complete the classification, one must also verify that
the combinations \(\gamma,\gamma_I,\gamma^t,\gamma^t_I\) close weakly
under the Poisson bracket and that the remaining constraints form a
non-degenerate second-class block.  The relevant brackets are
\begin{align}
\{\gamma(x),\gamma(y)\}&= 0, \nonumber \\
\{\gamma(x),\gamma_{J}(y)\}&=
-\varepsilon{_{J}}^{I}\gamma_{I}\delta(x-y)\approx0, \nonumber \\
\{\gamma_{I}(x),\gamma_{J}(y)\}&=
-2\varepsilon_{IJ}(\beta\Pi^{x}_{K}\gamma^{K}
+\alpha\Pi^{x}\gamma)\delta(x-y)\approx0, \nonumber \\
\{\gamma(x),\chi(y)\}&= 0, \nonumber \\
\{\gamma(x),\chi_{J}(y)\}&=
\varepsilon{^{I}}_{J}\chi_{I}\delta(x-y)\approx0, \nonumber \\
\{\gamma(x),\chi^{x}_{J}(y)\}&=
\varepsilon{^{I}}_{J}\chi^{x}_{I}\delta(x-y)\approx0, \nonumber \\
\{\gamma(x),\chi^{x}(y)\}&= 0, \nonumber \\
\{\gamma_{I}(x),\chi(y)\}&=
2\varepsilon_{IJ}\alpha(\varphi\chi^{J}
-e^{J}_{x}\chi^{x})\delta(x-y)\approx0, \nonumber \\
\{\gamma_{I}(x),\chi_{J}(y)\}&=
\varepsilon_{IK}(2\beta(e^{K}_{x}\chi^{x}_{J}
+\varphi_{J}\chi^{K})-\delta^{K}_{J}\chi)\delta(x-y)\approx0, \nonumber \\
\{\gamma_{I}(x),\chi^{x}_{J}(y)\}&=
-2\varepsilon_{IJ}(\beta\varphi^{K}\chi^{x}_{K}
+\alpha\varphi\chi^{x})\delta(x-y)\approx0, \nonumber \\
\{\gamma_{I}(x),\chi^{x}(y)\}&=
\varepsilon{_{I}}^{J}\chi^{x}_{J}\delta(x-y)\approx0, \nonumber \\
\{\chi_{I}(x),\chi^{x}_{J}(y)\}&= \eta_{IJ}\delta(x-y), \nonumber \\
\{\chi(x),\chi^{x}(y)\}&= \delta(x-y).
\label{28}
\end{align}
Equation~\eqref{28} defines a first-class constraint algebra with
field-dependent structure functions involving the canonical momenta.
Thus the algebra is a soft Hamiltonian gauge algebra in the standard
sense of constrained systems, rather than an ordinary finite-dimensional
Lie algebra with constant structure constants.  Its closure in
Eq.~\eqref{28} does not require the field equations.
This behavior closely mirrors the polynomial algebra of secondary
constraints in the Ashtekar formulation of canonical general
relativity~\cite{ashtekar1986new}, and similar quadratically nonlinear
structures have been observed in the Hamiltonian analysis of the
action~\eqref{1aa} by Grosse~\cite{grosse1992novel}.  A closely related
mechanism appears in the work of Ikeda and
Izawa~\cite{Ikeda1994_NonlinearGauge2D}.  In our case the algebra
reduces to the Poincar\'e gauge algebra $\mathrm{ISO}(1,1)$ when
$\alpha=\beta=\Lambda=0$.  Within the framework of \cite{ikeda1993quantum},
Yang--Mills theory can be viewed as a deformation of this topological
$\mathrm{ISO}(1,1)$ model; the deformed constraint algebra plays an
analogous role here.

Earlier Hamiltonian analyses of related forms of the KV model often work
in reduced variables or after imposing gauge conditions
\cite{Katanaev1988_TMP_Hamiltonian,strobl1993all,grignani1993canonical,%
kummer1993comment}.  The present analysis complements these treatments by
organizing the extended Dirac--Bergmann reduction explicitly, constructing
the associated Dirac brackets, and comparing them with the
Faddeev--Jackiw brackets in the same reduced variables.

The degree-of-freedom count displayed in the last row of
Table~\ref{tab:constraint-classification} refers only to local propagating
degrees of freedom.  The corresponding Casimir representative is derived
in Section~\ref{sec:casimir} from the reduced Dirac--Bergmann first-class
constraint ideal, after the reduced symplectic structure has been fixed by
the Dirac bracket construction and checked independently by the
Faddeev--Jackiw reduction.

\subsection{Dirac brackets}

Having separated the first- and second-class sectors, we now eliminate the
auxiliary second-class block while keeping the first-class gauge sector
intact.  The matrix of Poisson brackets among the second-class constraints
$\zeta^{\alpha}=(\chi,\chi_{I},\chi^{x}_{I},\chi^{x})$ is
\begin{align}
[C^{\alpha\beta}(x,y)]=
\left(
\begin{array}{cccc}
  0&0&0&\delta(x-y)\\
  0&0&\eta_{IJ}\delta(x-y)&0\\
  0&-\eta_{IJ}\delta(x-y)&0&0\\
  -\delta(x-y)&0&0&0\\
\end{array}
\right),
\label{29}
\end{align}
with inverse
\begin{align}
[C_{\alpha\beta}(x,y)]=
\left(
\begin{array}{cccc}
  0&0&0&-\delta(x-y)\\
  0&0&-\eta^{IJ}\delta(x-y)&0\\
  0&\eta^{IJ}\delta(x-y)&0&0\\
  \delta(x-y)&0&0&0\\
\end{array}
\right).
\label{30}
\end{align}
The Dirac bracket is
\begin{align}
\{A(x),B(y)\}_{D}=\{A(x),B(y)\}_{P}
-\int\!dudv\,\{A(x),\zeta^{\alpha}(u)\}
C_{\alpha\beta}(u,v)\{\zeta^{\beta}(v),B(y)\}.
\label{31}
\end{align}
Using~\eqref{30} and~\eqref{31}, the Dirac brackets of the theory are
\begin{align}
\{e^{I}_{\mu}(x),\Pi^{\nu}_{J}(y)\}_{D}&=
\delta^{I}_{J}\delta^{\nu}_{\mu}\delta(x-y),\nonumber \\
\{e^{I}_{x}(x),\varphi^{J}(y)\}_{D}&=
\eta^{IJ}\delta(x-y),\nonumber \\
\{\omega_{\mu}(x),\Pi^{\nu}(y)\}_{D}&=
\delta^{\nu}_{\mu}\delta(x-y),\nonumber \\
\{\omega_{x}(x),\varphi(y)\}_{D}&= \delta(x-y).
\label{32}
\end{align}
The spatial canonical pairs $(e^{I}_{x},\varphi^{J})$ and
$(\omega_{x},\varphi)$ acquire the nontrivial Dirac brackets displayed
in~\eqref{32}, a structure also found in the JT
model~\cite{cabrera2023canonical} and in two-dimensional quadratic
gravity formulated as a BF theory~\cite{cabrera2024hamiltonian}.

The Dirac brackets among the first-class constraints are
\begin{align}
\{\gamma(x),\gamma(y)\}_{D}&= 0,\nonumber \\
\{\gamma(x),\gamma_{J}\}_{D}&=
-\varepsilon{_{J}}^{K}\bigl(\gamma_{K}
+\varepsilon{_{K}}^{I}(2\beta\varphi^{N}\chi^{x}_{N}\chi_{I}
+2\alpha\varphi\chi_{I}\chi^{x}-\chi^{x}_{I}\chi)\bigr)\delta(x-y),
\nonumber \\
\{\gamma_{I}(x),\gamma_{J}(y)\}_{D}&=
2\varepsilon_{JI}\bigl((\beta\Pi^{x}_{K}\gamma^{K}
+\alpha\Pi^{x}\gamma)
+\mathcal{O}(\chi^{2},\chi\partial_x\chi)\bigr)\delta(x-y).
\label{33}
\end{align}
Comparing with the Poisson-bracket algebra~\eqref{28}, the Dirac
brackets provide a linear algebra for the first-class constraints in
which the contribution from the second-class sector survives only through
terms that are at least quadratic in the auxiliary second-class
constraints \cite{dirac1950generalized,amorim1995bft,%
henneaux1992quantization}.  In both descriptions the deformation is
controlled by $\alpha$, $\beta$, and $\Lambda$: when these are set to
zero the algebra reduces to the topological $\mathrm{ISO}(1,1)$ gauge
algebra.

The present work remains classical.  The reduced bracket structure is the
classical input underlying the standard BRST/BFV treatment of constrained
systems
\cite{becchi1976renormalization,fradkin1975quantization,fradkin1978quantization},
but no quantization is attempted here.  For the present analysis, the
bracket construction fixes the reduced symplectic structure needed to
reconstruct the extended Hamiltonian and the gauge generator in the same
constraint basis.

\subsection{The extended action and Hamiltonian}

We now reconstruct the extended action by determining the Lagrange
multipliers associated with the second-class sector.  They are fixed by
\begin{align}
\int dy\,\{\zeta^{\alpha}(x),H_{C}(y)\}
+\int dy\,\kappa^{\beta}(y)C^{\alpha\beta}(x,y)&\approx0, \nonumber \\
\kappa^{\alpha}&=-\int dy\,C_{\alpha\beta}(x,y)
\{\zeta^{\beta}(x),H_{C}(y)\},
\label{33aaaa}
\end{align}
where the inverse $C_{\alpha\beta}(x,y)$ satisfies
$\int dw\,C^{\alpha\beta}(x,w)C_{\beta\theta}(w,y)=\delta^{\alpha}_{\theta}\delta(x-y)$.
Using~\eqref{30} and~\eqref{33aaaa}, the Lagrange multipliers are
\begin{align}
\lambda&=\varepsilon{_{I}}^{J}\varphi_{J}e^{I}_{t}, \nonumber\\
\lambda^{I}_{x}&= D_{x}e^{I}_{t}
-\varepsilon{^{I}}_{J}e^{J}_{x}\omega_{t}
+2\beta\varepsilon_{JK}\varphi^{I}e^{J}_{t}e^{K}_{x}, \nonumber \\
\lambda_{x}&= \partial_{x}\omega_{t}
+2\alpha\varphi\varepsilon_{IJ}e^{I}_{t}e^{J}_{x}, \nonumber \\
\lambda^{I}&=-\varepsilon^{IJ}\varphi_{J}\omega_{t}
+\varepsilon{^{I}}_{J}e^{J}_{t}E.
\label{33cccc}
\end{align}
The extended action takes the form
\begin{align}
S_{E}=\int d^{2}x\,(\dot{Q}_{L}\Pi^{Q_{L}}
-\mathcal{H}-\tilde{\lambda}_{A}\Gamma_{A}-u_{A}\zeta_{A}),
\label{33a}
\end{align}
with
\begin{align}
\mathcal{H}=-\omega_{t}\gamma-e^{I}_{t}\gamma_{I},
\label{33b}
\end{align}
where $\tilde{\lambda}_{A}=(\tilde{\lambda},\tilde{\lambda}^{I},
\tilde{\lambda}^{I}_{t},\tilde{\lambda}_{t})$ and
$u_{A}=(u,u^{I},u^{I}_{x},u_{x})$ denote the Lagrange multipliers for
the first-class set $\Gamma_{A}=(\gamma,\gamma_{I},\gamma^{t}_{I},
\gamma^{t})$ and second-class set
$\zeta_{A}=(\chi,\chi_{I},\chi^{x}_{I},\chi^{x})$, respectively.  The
multipliers $u_{A}$ enforce $\zeta_{\alpha}\approx0$ without generating
gauge transformations, whereas $\tilde{\lambda}_{A}$ encode the gauge
freedom.  The extended Hamiltonian
\begin{align}
H_{E}=\int dx\,\bigl(\mathcal{H}
+\tilde{\lambda}\,\gamma
+\tilde{\lambda}^{I}\gamma_{I}
+\tilde{\lambda}^{I}_{t}\gamma^{t}_{I}
+\tilde{\lambda}_{t}\gamma^{t}\bigr)
\end{align}
is a linear combination of first-class constraints, ensuring that it
preserves the constraint surface.

\subsection{Gauge generator}

With the extended Hamiltonian written in the first-class basis, the gauge
transformations are obtained from the Castellani algorithm.  We define
\begin{align}
G=\int_{\Sigma}\bigl[D_{t}\tau^{I}_{t}\gamma^{t}_{I}
+D_{t}\tau_{t}\gamma^{t}
+\tau\gamma+\tau^{I}\gamma_{I}\bigr].
\label{34}
\end{align}
The gauge transformations on the extended phase space are
\begin{align}
\delta_{0}\varphi &= -\varepsilon{_{I}}^{J}\tau^{I}\varphi_{J},
\nonumber\\
\delta_{0}\varphi^{I} &=
\varepsilon^{IJ}(\varphi_{J}\tau-\tau_{J}E),\nonumber\\
\delta_{0}e^{I}_{0} &= D_{t}\tau^{I}_{t}, \nonumber \\
\delta_{0}e^{I}_{x} &=
\varepsilon{^{I}}_{K}e^{K}_{x}\tau-\partial_{x}\tau^{I}
-\varepsilon{^{I}}_{J}\omega_{x}\tau^{J}
-2\beta\varepsilon_{JK}e^{K}_{x}\varphi^{I}\tau^{J},\nonumber \\
\delta_{0}\omega_{t} &= D_{t}\tau_{t}, \nonumber \\
\delta_{0}\omega_{x} &=
-\partial_{x}\tau-2\alpha\varphi\varepsilon_{JK}e^{K}_{x}\tau^{J},
\nonumber\\
\delta_{0}\Pi &=
2\alpha\varepsilon_{JK}e^{K}_{x}\chi^{x}\tau^{J},\nonumber \\
\delta_{0}\Pi_{I} &=
\varepsilon{_{I}}^{J}\chi_{J}\tau
+(2\beta\varepsilon_{JK}(e^{K}_{x}\chi^{x}_{I}
-\chi^{K}\varphi_{I})-\varepsilon_{IJ}\chi)\tau^{J},\nonumber \\
\delta_{0}\Pi^{t}_{I} &= 0,\nonumber \\
\delta_{0}\Pi^{t} &= 0, \nonumber \\
\delta_{0}\Pi^{x}_{I} &=
\varepsilon{_{I}}^{J}(\varphi_{J}+\chi^{x}_{J})\tau
-\varepsilon_{IJ}(E+2\beta\varphi^{L}\chi^{x}_{L}
+2\alpha\varphi\chi^{x})\tau^{J},\nonumber \\
\delta_{0}\Pi^{x} &= -\varepsilon{_{I}}^{J}\varphi_{J}\tau^{I}.
\label{34a}
\end{align}
The transformations in~\eqref{34a} are interpreted within the extended
phase-space embedding.  The canonical momenta conjugate to the auxiliary
fields carry terms proportional to the auxiliary second-class constraints.
These terms do not represent additional physical gauge content; they
vanish once the second-class sector is eliminated and the natural
canonical pairs $(\omega_x,\varphi)$ and $(e^I_x,\varphi_I)$ are used.  In
this sense the present calculation complements earlier reduced treatments
\cite{ikeda1993quantum,strobl1993all}.

Choosing $\tau_t=-\tau=\zeta$ and $\tau^I_t=-\tau^I=\xi^I$, the gauge
symmetry takes the coordinate-independent form
\begin{align}
\varphi &\rightarrow&
\varphi+\varepsilon_{IJ}\xi^{I}\varphi^{J}, \nonumber \\
\varphi^{I} &\rightarrow&
\varphi^{I}+\varepsilon^{IJ}(-\varphi_{J}\zeta+\xi_{J}E), \nonumber \\
e^{I}_{\mu} &\rightarrow&
e^{I}_{\mu}+D_{\mu}\xi^{I}
+2\beta\varepsilon_{JK}e^{K}_{\mu}\xi^{J}\varphi^{I}
-\varepsilon{^{I}}_{J}e^{J}_{\mu}\zeta, \nonumber \\
\omega_{\mu} &\rightarrow&
\omega_{\mu}+\partial_{\mu}\zeta
+2\alpha\varphi\varepsilon_{JK}e^{K}_{\mu}\xi^{J}.
\label{34b}
\end{align}
These gauge symmetries reduce to Poincar\'e symmetries modulo terms
proportional to the equations of motion~\eqref{1bbb}, in agreement with
Refs.~\cite{ikeda1993quantum,strobl1993all,BlagojevicVukasinac1995_BRST}.
This closes the local Dirac--Bergmann analysis.  We next verify, by an
independent Faddeev--Jackiw reduction, that the same reduced symplectic
structure is obtained without separately classifying the constraints.


\section{Faddeev--Jackiw reduction}\label{sec:FJ}

The Faddeev--Jackiw (FJ) reduction provides an independent verification
of the reduced symplectic geometry obtained above.  Since the first-order KV
action already contains the symplectic potential displayed in
Eq.~\eqref{canonical-potential}, the FJ two-form should encode the same
canonical pairings, namely $(\omega_x,\varphi)$ and
$(e^I_x,\varphi_I)$, once the auxiliary second-class sector of the
extended Dirac formulation has been removed.  The section therefore
verifies that the FJ procedure yields the same reduced brackets and gauge
constraints as the Dirac--Bergmann analysis, without introducing
additional local dynamics.

We start from the first-order action~\eqref{15a}, written as
\begin{align}
\tilde{S}
=\int d^{2}x\big[\varphi\dot{\omega}_{x}
+\varphi_{I}\dot{e}^{I}_{x}
+\left(\partial_{x}\varphi+
\varepsilon{^{I}}_{J}e^{J}_{x}\varphi_{I}\right)\omega_{t}
+
\left(D_{x}\varphi_{I}\right)e^{I}_{t}-eE\big].
\label{41a}
\end{align}
Thus $\mathcal{L}^{(0)}=a^{(0)}_{a}(\xi)\dot\xi^{(0)a}-V^{(0)}$ with
\begin{align}
V^{(0)}&=-
\left(\partial_{x}\varphi+
\varepsilon{^{I}}_{J}e^{J}_{x}\varphi_{I}\right)\omega_{t}
-
\left(D_{x}\varphi_{I}\right)e^{I}_{t}
+eE.
\end{align}
The initial FJ variables and one-forms are chosen as
\begin{align}
\xi^{(0)a}=\{\varphi,\varphi_I,\omega_t,\omega_x,e^I_t,e^I_x\},
\qquad
 a^{(0)}_{a}=\{0,0,0,\varphi,0,\varphi_I\}.
\end{align}
The corresponding presymplectic matrix is
\begin{align}
f^{(0)}_{ab}(x,y)
&=
\frac{\delta a^{(0)}_{b}(y)}{\delta\xi^{(0)a}(x)}
-
\frac{\delta a^{(0)}_{a}(x)}{\delta\xi^{(0)b}(y)}
\nonumber \\
&=
\left(
\begin{array}{cccccc}
0&0&0&1&0&0\\
0&0&0&0&0&\delta^{I}_{J}\\
0&0&0&0&0&0\\
-1&0&0&0&0&0\\
0&0&0&0&0&0\\
0&-\delta^{J}_{I}&0&0&0&0
\end{array}
\right)\delta(x-y).
\label{43}
\end{align}
Here and below the matrices are displayed in block notation, with the
internal indices carried by the corresponding entries.  The matrix
$f^{(0)}_{ab}$ is singular.  Its independent zero modes along the
multiplier directions $\omega_t$ and $e^I_t$ give
\begin{align}
\Omega^{(0)} &= \partial_{x}\varphi
+\varepsilon{^{I}}_{J}e^{J}_{x}\varphi_{I}=0,
\nonumber \\
\Omega^{(0)}_{I} &= \partial_{x}\varphi_{I}
+\varepsilon{_{I}}^{J}\omega_{x}\varphi_{J}
-\varepsilon_{IJ}e^{J}_{x}E=0.
\label{45}
\end{align}
These are precisely the secondary constraints $\Psi$ and $\Psi_I$
obtained in the Dirac--Bergmann analysis after the auxiliary
second-class sector has been eliminated.  In the FJ language, the next
step is to test whether the zero modes of the enlarged matrix generate
additional constraints.  The explicit matrix and its zero modes are
reported in Appendix~\ref{app:FJ-matrices}.  Their contraction with the
corresponding vector $Z_c$ vanishes identically; hence no further
constraints arise.  The residual degeneracy of the next presymplectic
matrix is therefore gauge degeneracy rather than a signal of new constraints.

The constraints in Eq.~\eqref{45} are incorporated into the symplectic
Lagrangian by replacing the original multipliers with velocities,
$\omega_t=\dot\lambda$ and $e^I_t=\dot\lambda^I$.  This gives
\begin{equation}
\mathcal{L}^{(1)}=
\varphi\dot{\omega}_{x}
+\varphi_{I}\dot{e}^{I}_{x}
+\Omega^{(0)}_{I}\dot{\lambda}^{I}
+\Omega^{(0)}\dot{\lambda}-V^{(1)},
\label{50}
\end{equation}
with
\begin{equation}
V^{(1)}=V^{(0)}\big|_{\Omega^{(0)}=0,\,\Omega^{(0)}_I=0}=0.
\end{equation}
Thus the volume Hamiltonian disappears on the constraint surface, as
expected for a generally covariant two-dimensional gauge system.  The
vanishing of $V^{(1)}$ is understood in this restricted FJ sense: after
the constraints have been absorbed into the kinetic part of the
first-order Lagrangian, the remaining potential is zero.

The matrix built from~\eqref{50} is still singular.  Since no additional
constraints are produced, one may introduce gauge-fixing conditions in
order to invert the final presymplectic form and extract the generalized
FJ brackets.  We use the temporal gauge for the original multiplier
sector,
\begin{align}
\omega_t=0,
\qquad
 e^I_t=0,
\label{52}
\end{align}
and implement it in the FJ matrix through the corresponding multiplier
variables.  This gauge fixing serves only to invert the presymplectic matrix.  It
does not enter the gauge-unfixed Dirac construction of the Casimir in
Sec.~\ref{sec:casimir}, nor does it modify the reduced first-class
constraint ideal obtained after removing the auxiliary second-class
sector.

With the gauge-fixing multipliers included, the final matrix
$f^{(2)}_{ab}$ is invertible.  The explicit form of $f^{(2)}_{ab}$ and
of the entries of its inverse relevant for the reduced brackets is given
in Appendix~\ref{app:FJ-matrices}.  The generalized FJ brackets are
\begin{align}
\{\omega_{x}(x),\varphi(y)\}_{FJ}&=\delta(x-y),
\nonumber \\
\{e^{J}_{x}(x),\varphi_{I}(y)\}_{FJ}&=
\delta^{J}_{I}\delta(x-y).
\label{56}
\end{align}
These are exactly the Dirac brackets~\eqref{32} after the identifications
\begin{equation}
\Pi^x=\varphi,
\qquad
\Pi^x_I=\varphi_I
\end{equation}
are imposed strongly by the second-class sector.  Equivalently,
\begin{align}
\{\omega_x(x),\varphi(y)\}_{FJ}
&=
\{\omega_x(x),\Pi^x(y)\}_{D}\big|_{\Pi^x=\varphi},
\nonumber\\
\{e^I_x(x),\varphi_J(y)\}_{FJ}
&=
\{e^I_x(x),\Pi^x_J(y)\}_{D}\big|_{\Pi^x_J=\varphi_J}.
\end{align}
The agreement is structural: both procedures lead to the same reduced
symplectic geometry.

The local degree-of-freedom count is also the same.  The reduced FJ
variables are $(\omega_x,\varphi)$ and $(e^I_x,\varphi_I)$, giving six
phase-space variables.  The zero-mode analysis produces the three gauge
constraints $\Omega^{(0)}$ and $\Omega^{(0)}_I$, which correspond to
first-class constraints in the Dirac description.  Hence
\begin{equation}
N_{\rm phys}
=
\frac12\left(6-2\times3\right)=0.
\label{FJ-dof-count}
\end{equation}
The gauge-fixing conditions used to invert $f^{(2)}_{ab}$ do not enter as
additional physical restrictions in this count.  They remove the
presymplectic degeneracy so that the generalized brackets can be read
explicitly.

\subsection{Gauge generator in the MW method}\label{sec:eq}

The Montani--Wotzasek (MW) version of the FJ formalism~\cite{Mon}
provides a direct route to the gauge transformations from the zero modes
of the presymplectic matrix.  For the present model, the relevant zero modes of
the matrix associated with~\eqref{50} may be written as
\begin{align}
(w^{(1)})_{1}^{T}
&=\bigl(0,-\varepsilon{_{I}}^{K}\varphi_{K}\delta(x-y),
-\partial_{x}\delta(x-y),
-\varepsilon{^{I}}_{K}e^{K}_{x}\delta(x-y),0\bigr),
\nonumber\\
(w^{(1)})_{2}^{T}
&=\bigl(\varepsilon{_{N}}^{J}\varphi_{J},
-\varepsilon_{NI}E\delta(x-y),
2\alpha\varphi\varepsilon_{NJ}e^{J}_{x}\delta(x-y),
\nonumber\\
&\quad -\delta^{I}_{N}\partial_{x}\delta(x-y)
-\varepsilon{_{N}}^{I}\omega_{x}\delta(x-y)
+2\beta\varepsilon_{NL}e^{L}_{x}\varphi^{I}\delta(x-y),
0,\delta(x-y)\bigr).
\label{58a}
\end{align}
According to the MW prescription, multiplying the functional variation
of $\mathcal{L}^{(1)}$ by these zero modes gives the gauge
transformations over the full configuration space.  With the parameter
identification used in the Dirac analysis, one obtains
\begin{align}
\delta_{0}\varphi &= -\varepsilon{_{I}}^{J}\tau^{I}\varphi_{J},
\nonumber\\
\delta_{0}\varphi^{I} &=
\varepsilon^{IJ}(\varphi_{J}\tau-\tau_{J}E),
\nonumber\\
\delta_{0}e^{I}_{t} &= D_{t}\tau^{I},
\nonumber \\
\delta_{0}e^{I}_{x} &=
\varepsilon{^{I}}_{K}e^{K}_{x}\tau
-\partial_{x}\tau^{I}
-\varepsilon{^{I}}_{J}\omega_{x}\tau^{J}
-2\beta\varepsilon_{JK}e^{K}_{x}\varphi^{I}\tau^{J},
\nonumber \\
\delta_{0}\omega_{t} &= \partial_{t}\tau,
\nonumber \\
\delta_{0}\omega_{x} &=
-\partial_{x}\tau-2\alpha\varphi\varepsilon_{JK}e^{K}_{x}\tau^{J}.
\label{60a}
\end{align}
These transformations agree with the Dirac result~\eqref{34b} once the
same convention for the multiplier sector is adopted; the relative sign
of the $\partial_t\tau$ term in $\delta_0\omega_t$ reflects the opposite
parameter identification $\tau_t=-\tau$ used in Eq.~\eqref{34b}.  The
FJ/MW construction supplies a symplectic counterpart of the Dirac
constraint analysis.  It reproduces the reduced brackets, the gauge
constraints and the local degree-of-freedom count, while avoiding an
independent classification of constraints into first- and second-class
subsets.  This agrees with the general correspondence between the
symplectic and Dirac reductions of first-order Lagrangians
\cite{BarcelosWotzasek1992,Mon}.

This completes the symplectic check of the reduced Dirac structure.  The
next step uses this common reduced phase space in a different way: the
Casimir is not obtained from the gauge-fixed Faddeev--Jackiw matrix, but
from the gauge-unfixed Dirac--Bergmann first-class constraint ideal.


\section{Gauge-unfixed Casimir from the reduced constraint ideal}\label{sec:casimir}

The Faddeev--Jackiw analysis of the previous section verifies the
reduced symplectic structure, but the Casimir itself is controlled by the
reduced Dirac--Bergmann first-class constraint ideal.  The two reductions
therefore play distinct roles.  The Dirac--Bergmann analysis identifies
the first-class generators whose ideal controls spatial conservation,
whereas the Faddeev--Jackiw reduction verifies that the same reduced
brackets and canonical pairs are obtained without relying on the extended
phase-space embedding.  The Casimir derived below is therefore a
gauge-unfixed object of the common reduced phase space.

The local constraint classification fixes the propagating content,
whereas the reduced phase space also contains a global invariant: the
Katanaev/Poisson--Sigma Casimir.  This conserved quantity plays a central
role in the integrability of the field equations
\cite{katanaev1990complete,strobl1993all,katanaev2002effective,%
GrumillerKummerVassilevich2002}.  A representative of this Casimir is
obtained directly from the gauge-unfixed constraint ideal, without
imposing the conformal gauge used in several earlier Hamiltonian
treatments.

After the auxiliary second-class sector is eliminated by the Dirac
bracket, or equivalently after imposing
\begin{equation}
\chi=\chi_I=\chi^x=\chi_I^x=0
\label{second-class-strong-casimir}
\end{equation}
as strong equalities, the first-class constraints~\eqref{26} reduce to
the secondary constraints
\begin{align}
\Psi &= \partial_x\varphi
+\varepsilon^I{}_{J}e^J_x\varphi_I\approx0,
\label{reduced-Psi-Casimir}\\
\Psi_I &= \partial_x\varphi_I
+\varepsilon_I{}^J\omega_x\varphi_J
-\varepsilon_{IJ}e^J_xE\approx0,
\label{reduced-PsiI-Casimir}
\end{align}
where $E=\alpha\varphi^2+\beta\varphi_I\varphi^I+\Lambda$.  This is a
reduction of the auxiliary second-class sector, not a gauge fixing: it
removes only the redundancy introduced by treating $\varphi$ and
$\varphi_I$ as configuration variables.  No condition has
been imposed on $e^I_x$, $\omega_x$, $e^I_t$, $\omega_t$, or on the
spacetime coordinate freedom.

We look for a Lorentz-invariant representative of the Casimir in the
form
\begin{equation}
\mathcal{C}=\mathcal{C}(\varphi,\rho),\qquad
\rho:=\varphi_I\varphi^I.
\label{Casimir-ansatz}
\end{equation}
This is the natural scalar ansatz built from the target-space variables
of the first-order KV action.  More general representatives are obtained
by applying a regular reparametrization $\mathcal{C}\mapsto F(\mathcal{C})$.
The spatial derivative of $\mathcal{C}$ is
\begin{equation}
\partial_x\mathcal{C}
=\mathcal{C}_{\varphi}\partial_x\varphi
+2\mathcal{C}_{\rho}\varphi^I\partial_x\varphi_I,
\label{Casimir-derivative-1}
\end{equation}
where $\mathcal{C}_{\varphi}:=\partial\mathcal{C}/\partial\varphi$ and
$\mathcal{C}_{\rho}:=\partial\mathcal{C}/\partial\rho$.  Solving
\eqref{reduced-Psi-Casimir} and \eqref{reduced-PsiI-Casimir} for the
spatial derivatives gives
\begin{align}
\partial_x\varphi &= \Psi-\varepsilon^I{}_{J}e^J_x\varphi_I,
\label{dphi-from-Psi}\\
\partial_x\varphi_I &= \Psi_I-\varepsilon_I{}^J\omega_x\varphi_J
+\varepsilon_{IJ}e^J_xE.
\label{dphiI-from-PsiI}
\end{align}
Substitution into \eqref{Casimir-derivative-1} yields
\begin{align}
\partial_x\mathcal{C}
&= \mathcal{C}_{\varphi}\Psi
+2\mathcal{C}_{\rho}\varphi^I\Psi_I
-\mathcal{C}_{\varphi}\varepsilon^I{}_{J}e^J_x\varphi_I
-2\mathcal{C}_{\rho}\varphi^I\varepsilon_I{}^J\omega_x\varphi_J
+2\mathcal{C}_{\rho}E\varphi^I\varepsilon_{IJ}e^J_x.
\label{Casimir-derivative-expanded}
\end{align}
The term proportional to $\omega_x$ vanishes because $\varphi^I\varphi_J$
is symmetric in internal indices whereas $\varepsilon_I{}^J$ is
antisymmetric.  Using $\varepsilon^I{}_{J}\varphi_I=\varepsilon_{IJ}\varphi^I$,
the remaining non-constraint term is proportional to $e^J_x$, giving
\begin{equation}
\partial_x\mathcal{C}
=\mathcal{C}_{\varphi}\Psi
+2\mathcal{C}_{\rho}\varphi^I\Psi_I
+\left(2E\mathcal{C}_{\rho}-\mathcal{C}_{\varphi}\right)
\varepsilon_{IJ}\varphi^Ie^J_x.
\label{Casimir-derivative-precondition}
\end{equation}
Because $e^J_x$ remains arbitrary, the last term must vanish as an
identity.  The Casimir condition is therefore
\begin{equation}
\boxed{\mathcal{C}_{\varphi}=2E\mathcal{C}_{\rho},}
\label{Casimir-PDE}
\end{equation}
or explicitly
\begin{equation}
\frac{\partial\mathcal{C}}{\partial\varphi}
=2\left(\alpha\varphi^2+\beta\rho+\Lambda\right)
\frac{\partial\mathcal{C}}{\partial\rho}.
\label{Casimir-PDE-explicit}
\end{equation}
Equation~\eqref{Casimir-PDE-explicit} is obtained entirely from the
reduced first-class constraint ideal; no conformal or light-cone gauge
has been used.  It coincides with the Poisson--Sigma Casimir condition
$P^{ij}\partial_j C=0$ written in the present variables
\cite{SchallerStrobl1994,GrumillerKummerVassilevich2002}; the derivation
above shows that the same condition is obtained off shell inside the
extended Dirac--Bergmann reduction, without using the target-space
Poisson geometry as an input.

Solving \eqref{Casimir-PDE-explicit} by characteristics gives
\begin{equation}
\frac{d\rho}{d\varphi}+2\beta\rho=-2(\alpha\varphi^2+\Lambda).
\label{Casimir-characteristic-linear}
\end{equation}
The integrating factor is $e^{2\beta\varphi}$, so a convenient
representative of the invariant is
\begin{equation}
\boxed{
\mathcal{C}_{\rm DB}
=e^{2\beta\varphi}\varphi_I\varphi^I
+2\int^{\varphi}ds\,(\alpha s^2+\Lambda)e^{2\beta s}
}
\label{Casimir-DB}
\end{equation}
up to an irrelevant additive constant and regular reparametrizations
$F(\mathcal{C}_{\rm DB})$.  The verification follows from
\begin{equation}
(\mathcal{C}_{\rm DB})_{\rho}=e^{2\beta\varphi},
\qquad
(\mathcal{C}_{\rm DB})_{\varphi}=2E\,e^{2\beta\varphi},
\label{Casimir-partials}
\end{equation}
which satisfy~\eqref{Casimir-PDE} identically and give
\begin{equation}
\boxed{
\partial_x\mathcal{C}_{\rm DB}
=2e^{2\beta\varphi}\left(E\Psi+\varphi^I\Psi_I\right),
}
\label{Casimir-derivative-final}
\end{equation}
so that $\partial_x\mathcal{C}_{\rm DB}\approx0$ on the constraint
surface; $\mathcal{C}_{\rm DB}$ is therefore constant along each
connected component of $\Sigma$. Having obtained the invariant directly
from the reduced Dirac--Bergmann constraint ideal, we now identify it with
the standard integrability invariant of the KV model and with the
corresponding Poisson--Sigma Casimir.

This comparison does not introduce a new conserved quantity.  Rather, it
shows that the gauge-unfixed canonical representative found above
reproduces the known invariant.  The conserved quantity obtained here
coincides, up to the standard normalization of the light-cone variables,
with the integral of motion of the Katanaev--Volovich model
\cite[Eqs.~(71),~(92),~(93)]{katanaev2002effective}, also found
independently in two-dimensional dilaton gravity
\cite{GegenbergKunstatterLouisMartinez1995}. In Katanaev's notation this
integral is
\begin{equation}
A_{\rm K}
=
p^+p^-e^{-Q(\pi)}-W(\pi),
\qquad
Q'(\pi)=U(\pi),
\qquad
W'(\pi)=V(\pi)e^{-Q(\pi)} .
\label{Katanaev-integral-motion}
\end{equation}
The use of light-cone components is only a change of basis in the
internal Lorentz space, not a spacetime gauge choice.  With
$p^{\pm}=(p^0\pm p^1)/\sqrt2$, one has
\begin{equation}
p^+p^-=\frac{1}{2} p_a p^a .
\label{light-cone-contraction}
\end{equation}
The dictionary with the variables and density conventions used here is
\begin{equation}
\pi_{\rm K}=-\varphi,
\qquad
p_a p^a\leftrightarrow \varphi_I\varphi^I,
\qquad
U_{\rm K}=2\beta,
\qquad
V_{\rm K}(\pi_{\rm K})=\alpha\pi_{\rm K}^2+\Lambda .
\label{Katanaev-DB-dictionary}
\end{equation}
The minus sign in $\pi_{\rm K}=-\varphi$ reflects the relative
curvature-density and orientation conventions between Katanaev's
first-order action and the conventions adopted in~\eqref{1c}.
Consequently,
\begin{equation}
Q(\pi_{\rm K})=2\beta\pi_{\rm K}=-2\beta\varphi,
\qquad
e^{-Q}=e^{2\beta\varphi},
\label{Katanaev-Q-map}
\end{equation}
and, after the change of integration variable $s=-u$ and after fixing
irrelevant additive constants in the primitives,
\begin{equation}
W(\pi_{\rm K})
=
\int^{\pi_{\rm K}}ds\,
(\alpha s^2+\Lambda)e^{-2\beta s}
=
-w_\beta(\varphi),
\qquad
w_\beta(\varphi):=\int^{\varphi}du\,
(\alpha u^2+\Lambda)e^{2\beta u} .
\label{Katanaev-W-map}
\end{equation}
Therefore
\begin{equation}
A_{\rm K}
=
\frac12 e^{2\beta\varphi}\varphi_I\varphi^I
+w_\beta(\varphi),
\qquad
\mathcal C_{\rm DB}
=
e^{2\beta\varphi}\varphi_I\varphi^I+2w_\beta(\varphi)
=2A_{\rm K} .
\label{Casimir-Katanaev-relation}
\end{equation}
In the static normalization used below,
$\mathcal C_{\rm DB}=-2M_{\rm Cas}$, and hence
$A_{\rm K}=-M_{\rm Cas}$.  Thus the difference between the two conserved
quantities is only the conventional normalization of the light-cone
product and the sign convention relating Katanaev's dilaton to the scalar
variable used here.

The two derivations nevertheless differ in structure.  In
\cite{katanaev2002effective}, the invariant is obtained \emph{a posteriori}
as an integral of motion of the field equations; its conservation is
established on shell, and its first-class and central character is then
verified afterwards.  Here, by contrast, $\mathcal C_{\rm DB}$ is
constructed \emph{off shell} from the reduced first-class constraint ideal:
requiring that $\partial_x\mathcal C$ lie in the ideal for arbitrary
$e^J_x$ yields the defining condition~\eqref{Casimir-PDE}, from which
$\mathcal C_{\rm DB}$ follows without integrating the field equations or
fixing a gauge.  The ideal membership~\eqref{Casimir-derivative-final} is
therefore a derived identity rather than an \emph{a posteriori} check.
Thus the Casimir conservation underlying the integrability analysis is
recovered here directly from the canonical constraint structure.

\begin{proposition}
After the auxiliary second-class sector has been imposed strongly, the
Katanaev/Poisson--Sigma Casimir admits a gauge-unfixed canonical
representative whose spatial derivative belongs to the reduced
first-class constraint ideal.  In the normalization used here this
representative is $\mathcal C_{\rm DB}$ in Eq.~\eqref{Casimir-DB}, and
Eq.~\eqref{Casimir-derivative-final} gives the explicit ideal
membership.
\end{proposition}

In terms of the full extended phase space, the Casimir may equivalently
be represented on the second-class surface as
\begin{equation}
\mathcal{C}_{\rm DB}
\doteq
e^{2\beta\Pi^x}\Pi^x_I\Pi^{xI}
+2\int^{\Pi^x}ds\,(\alpha s^2+\Lambda)e^{2\beta s},
\label{Casimir-canonical-momenta}
\end{equation}
where $\doteq$ denotes equality after imposing the second-class
constraints.  From~\eqref{26}, one has
$\gamma=\Psi+\mathcal{O}(\chi,\partial_x\chi^x)$ and
$\gamma_I=\Psi_I+\mathcal{O}(\chi,\chi^x,\partial_x\chi^x_I)$, so
\begin{equation}
\partial_x\mathcal{C}_{\rm DB}
\in
\langle\gamma,\gamma_I,\chi,\chi_I,\chi^x,\chi_I^x\rangle.
\label{Casimir-ideal-full}
\end{equation}
This gives the canonical sense in which the Katanaev invariant is
contained in the present Dirac--Bergmann formulation.

For $\beta\neq0$, evaluating the integral in~\eqref{Casimir-DB} by
successive integration by parts gives the closed form
\begin{equation}
\mathcal{C}_{\rm DB}
=e^{2\beta\varphi}\!\left[
\varphi_I\varphi^I
+\frac{\alpha}{\beta}\varphi^2
-\frac{\alpha}{\beta^2}\varphi
+\frac{\alpha}{2\beta^3}
+\frac{\Lambda}{\beta}
\right]+\mathrm{const},
\label{Casimir-DB-explicit}
\end{equation}
while the $\beta=0$ case gives
\begin{equation}
\mathcal{C}_{\rm DB}\big|_{\beta=0}
=\varphi_I\varphi^I+\frac{2\alpha}{3}\varphi^3+2\Lambda\varphi.
\label{Casimir-beta-zero}
\end{equation}
These expressions agree with the standard Poisson--Sigma/dilaton-gravity
Casimir $C_{\rm PSM}=e^{Q}Y+w$ of \cite{GrumillerKummerVassilevich2002},
with the identifications $X\leftrightarrow\varphi$,
$Y\leftrightarrow\rho/2$, $U=2\beta$, and $V=\alpha X^2+\Lambda$:
specifically $\mathcal{C}_{\rm DB}=2C_{\rm PSM}$, where the factor of
two reflects the normalization convention $\rho=\varphi_I\varphi^I$
versus $Y=\rho/2$.  These identifications refer to the normalization of
the action~\eqref{1c}; other conventions for the KV model may distribute
signs and numerical factors differently.  The overall normalization is
physically irrelevant since any regular function of a Casimir is again a
Casimir.

Fixing a value
$\mathcal{C}_{\rm DB}=\mathcal{C}_0$ selects a global sector of classical
solutions, while the local degree-of-freedom count in
Table~\ref{tab:constraint-classification} remains unchanged.  The
construction so far is local on the spatial slice and identifies the
Casimir as a reduced canonical invariant.  We now use a fixed value of this
invariant to select a static torsionful branch and to determine how the
Casimir normalization enters the metric functions and the horizon
thermodynamics.


\section{Static torsionful sector and Casimir-normalized thermodynamics}
\label{sec:BH}

The canonical analysis of Sections~\ref{sec:Dirac}--\ref{sec:casimir}
is local on the spatial slice.  It determines the reduced symplectic
structure, the first-class constraint ideal and the gauge-unfixed Casimir
representative without selecting a particular global sector.  This
canonical information is then used to construct a static torsionful branch
of the first-order KV model.  In this branch, the Casimir normalization
fixes the radial metric field, whereas the metric Killing norm contains an
additional torsional factor.  This separation controls both the horizon
temperature and the first-law normalization.

Throughout this section we set
\begin{equation}
X:=\varphi,\qquad \rho:=\varphi_I\varphi^I,
\qquad
V(X):=\alpha X^2+\Lambda,
\label{BH-XrhoV}
\end{equation}
and use
\begin{equation}
Q(X)=2\beta X,
\qquad
w_\beta(X)=\int_0^X ds\,e^{2\beta s}(\alpha s^2+\Lambda)
          =\int_0^X ds\,e^{Q(s)}V(s).
\label{BH-Qw-def}
\end{equation}
With these definitions, the canonical Casimir~\eqref{Casimir-DB} becomes
\begin{equation}
\mathcal C_{\rm DB}=e^{Q(X)}\rho+2w_\beta(X).
\label{BH-Casimir-general}
\end{equation}
On every connected solution the Casimir takes a constant value,
\(\mathcal C_{\rm DB}=\mathcal C_0\).  In this section we use the
Casimir normalization
\begin{equation}
\boxed{M_{\rm Cas}:=-\frac12\mathcal C_0}
\label{BH-mass-normalization}
\end{equation}
and choose the reference background so that \(M_{\rm Cas}=0\) when
\(\mathcal C_0=0\).  This normalization defines the Hamiltonian parameter
of the static branch; assigning a universal ADM-type mass requires
boundary data and a choice of time normalization.

The functions \(Q\) and \(w_\beta\) in Eq.~\eqref{BH-Qw-def} are the
specialization of the standard Poisson--Sigma/dilaton-gravity functions
to the KV potential.  The present analysis reconstructs these standard
functions from the gauge-unfixed Dirac--Bergmann Casimir and fixes their
role as Casimir-normalized Hamiltonian parameters in the KV variables.
It fixes the Hamiltonian normalization used below, while the entropy is
treated through the standard first-order dilaton-gravity horizon formula.

\subsection{Static torsionful sector in the Casimir-normalized frame}
\label{subsec:torsionful-static}

We work in dilaton gauge,
\begin{equation}
X=\varphi=r,
\label{BH-dilaton-gauge-general}
\end{equation}
and take a \emph{general} static diagonal representative
\begin{equation}
ds^2=N(r)\,dt^2-B(r)\,dr^2,
\qquad
e^0{}_t=\sqrt{N},\quad e^1{}_r=\sqrt{B},
\label{BH-KV-metric}
\end{equation}
without imposing the Schwarzschild relation $NB=1$, which is not preserved
by the torsionful branch for $\beta\neq0$.  The Lorentz-scalar \(\rho=\varphi_I\varphi^I\) is fixed by the
connection equation together with the dilaton gauge.  The equation
obtained by varying the Lorentz connection implies
\(\partial_\mu X+\varepsilon^{IJ}\varphi_I e_{\mu J}=0\); contracting it
with itself gives the invariant identity
\begin{equation}
\rho=-g^{\mu\nu}\partial_\mu X\partial_\nu X=g^{rr}\cdot(-1)=\frac{1}{B}.
\label{BH-rho-gradient}
\end{equation}
This relation fixes the \emph{radial} metric, \(B=1/\rho\), but not the
Killing norm \(N=g_{tt}\), which is determined below by the remaining
field equations.  Writing the on-shell value of \(\rho\) as
the variable \(\xi(r):=\rho(r)\) and substituting into the Casimir relation
\eqref{BH-Casimir-general} with the normalization
\eqref{BH-mass-normalization},
\begin{equation}
e^{2\beta r}\xi(r)+2w_\beta(r)=-2M_{\rm Cas}
\;\Longrightarrow\;
\boxed{\ \xi(r)=\rho(r)=e^{-2\beta r}\left[-2M_{\rm Cas}-2w_\beta(r)\right].\ }
\label{BH-xi-beta}
\end{equation}
This is the Casimir-normalized radial field, carrying \(e^{-2\beta r}\),
and it is distinct from the Killing norm.  Solving the remaining equations
in~\eqref{1bbb} fixes the rest of the configuration.  With \(B=1/\xi\),
the field equations are satisfied identically in $r$ by
\begin{equation}
\boxed{\ N(r)=e^{4\beta r}\xi(r)=-2e^{2\beta r}\bigl(w_\beta(r)+M_{\rm Cas}\bigr),
\qquad NB=e^{4\beta r},\ }
\label{BH-Killing-norm}
\end{equation}
as verified by direct substitution into the $\varphi$-, $\varphi_I$-,
$e^I$- and $\omega$-equations.  The corresponding non-vanishing components
are \(e^0{}_t=e^{2\beta r}\sqrt{\xi}\), \(e^1{}_r=\xi^{-1/2}\),
\(\omega=\omega_t\,dt\) with \(\omega_t=-e^{2\beta r}E\),
\((\varphi_0)^2=\xi\), \(\varphi_1=0\), and
\(E=\alpha r^2+\beta\xi+\Lambda\).  The metric functions read off from the
zweibein are \(N=(e^0{}_t)^2=e^{4\beta r}\xi\) and
\(B=(e^1{}_r)^2=\xi^{-1}\), so that the volume is
\(e=\det e^I_\mu=e^{2\beta r}\), not unity, and \(NB=e^{4\beta r}\).
The geometry is therefore not in Schwarzschild gauge for \(\beta\neq0\):
the Killing norm \(N=e^{4\beta r}\xi\) and the Casimir-normalized
radial field \(\xi\) coincide only in the torsionless limit
\(\beta=0\), where \(N=\xi\) and \(NB=1\).
For \(\beta=0\), Eq.~\eqref{BH-Killing-norm} reduces to the torsionless
expression
\begin{equation}
N(r)=\xi(r)=-2M_{\rm Cas}-2\left(\frac{\alpha r^3}{3}+\Lambda r\right),
\label{BH-xi-beta-zero-limit}
\end{equation}
which is the sector obtained when \(\varphi_I\) imposes \(T^I=0\) in the
first-order action.  The limiting statement should be understood at the
level of the first-order action~\eqref{1c}, not as a direct substitution
\(\beta=0\) in the second-order action~\eqref{1aa}, where the
coefficient of the torsion-squared term is written as \(1/\beta\).

The sector is genuinely torsional for \(\beta\neq0\).  The field equation
obtained by varying \(\varphi_I\) is
\begin{equation}
T^I-2\beta e\,\varphi^I=0.
\label{BH-torsion-eq-general}
\end{equation}
We distinguish the density \(T^I\) from the scalar normalization obtained
by dividing by the zweibein determinant.  We define
\begin{equation}
\mathcal T^I:=e^{-1}T^I .
\label{BH-normalized-torsion-def}
\end{equation}
Then the torsion equation implies
\begin{equation}
\mathcal T_I\mathcal T^I
=4\beta^2\varphi_I\varphi^I=4\beta^2\rho=4\beta^2\xi(r).
\label{BH-torsion-invariant}
\end{equation}
Equivalently, for the density itself,
\(T_I T^I=4\beta^2 e^2\rho=4\beta^2 e^{4\beta r}\xi\) in the
diagonal representative.  The normalized torsion scalar is non-zero in
the static region whenever \(\beta\neq0\) and \(\xi>0\), while it
vanishes at Killing horizons where \(\xi=0\).  This statement is
independent of the local Lorentz boost used to represent the zweibein.

\subsection{Poisson--Sigma and invariant on-shell checks}
\label{subsec:torsionful-onshell-checks}

The previous subsection used the constant value of
\(\mathcal C_{\rm DB}\) to construct an explicit static torsionful branch
in a diagonal Casimir-normalized frame.  The Poisson--Sigma identification
of the Casimir has already fixed the normalization in
Sec.~\ref{sec:casimir}; here it is used only as an invariant check of the
static representative constructed above.  We verify that the scalar
relations used above--in particular
\(\rho=-g^{\mu\nu}\partial_\mu X\partial_\nu X\), radial conservation of
the Casimir and the normalized torsion invariant--follow from the
covariant first-order field equations rather than from the diagonal
representative alone.

The first-order KV model is a Poisson--Sigma model with target variables
\((X,\varphi_I)\) and connection one-forms \((\omega,e^I)\).  In the
present conventions its Casimir is
\begin{equation}
\mathcal C_{\rm DB}=e^{Q(X)}\rho+2w_\beta(X),
\qquad
Q(X)=2\beta X,
\qquad
w_\beta'(X)=e^{Q(X)}V(X),
\label{BH-PSM-Casimir-check}
\end{equation}
which is \(2C_{\rm PSM}\) in the standard notation
\(C_{\rm PSM}=e^{Q}Y+w\), with \(Y=\rho/2\)
\cite{GrumillerKummerVassilevich2002}.  The general PSM field equations
imply that this Casimir is constant on every connected solution.  Hence
using \(\mathcal C_{\rm DB}=\mathcal C_0\) and the dilaton gauge \(X=r\)
gives the static relation \eqref{BH-xi-beta}.  This identifies the Casimir scalar field \eqref{BH-xi-beta} as the KV
specialization of the general PSM on-shell relation, written in the
Casimir-normalized diagonal representative.

Several invariant checks follow directly.  First, the connection
equation in \eqref{1bbb} implies
\begin{equation}
\partial_\mu X+\varepsilon^{IJ}\varphi_Ie_{\mu J}=0.
\label{BH-omega-eom-static-check}
\end{equation}
Contracting this equation with itself gives
\begin{equation}
\rho=-g^{\mu\nu}\partial_\mu X\partial_\nu X.
\label{BH-rho-gradient-check}
\end{equation}
For \(X=r\) and \(g_{rr}=-1/\xi\), this reproduces
\(\rho=\xi\) (the relation fixes the \emph{radial} metric, $B=1/\xi$, not
the Killing norm).  Second, the reduced constraint identity
\eqref{Casimir-derivative-final} gives
\(\partial_r\mathcal C_{\rm DB}=0\) on the first-class constraint surface.
Using \(\rho=\xi\), this is
\begin{equation}
\frac{d}{dr}\left(e^{2\beta r}\xi+2w_\beta(r)\right)=0,
\label{BH-onshell-C-constant}
\end{equation}
which is equivalent to
\begin{equation}
\xi'(r)=-2\left[V(r)+\beta\xi(r)\right]
          =-2\left(\alpha r^2+\Lambda+\beta\xi(r)\right).
\label{BH-xi-prime-constraint-form}
\end{equation}
Equation~\eqref{BH-xi-beta} is the integrated form of this first-order
constraint equation with integration constant \(-2M_{\rm Cas}\).  Finally,
the \(\varphi_I\)-equation in \eqref{1bbb} gives
\begin{equation}
T^I=2\beta e\,\varphi^I,
\label{BH-torsion-eom-check}
\end{equation}
and therefore, in the diagonal representative where \(e=e^{2\beta r}\),
\begin{equation}
\mathcal T_I\mathcal T^I=4\beta^2\rho=4\beta^2\xi,
\qquad \mathcal T^I:=e^{-1}T^I.
\label{BH-torsion-invariant-check}
\end{equation}
For the unnormalised density one instead has
\(T_IT^I=4\beta^2e^2\rho=4\beta^2e^{4\beta r}\xi\).
These statements verify the static sector at the level of Lorentz-scalar
on-shell relations and of the PSM equations, and the explicit
configuration of Section~\ref{subsec:torsionful-static} satisfies all four
field equations; the independent residuals are displayed in
Appendix~\ref{app:torsionful-residuals}.  These checks do not select a
unique boost representative of \(e^I{}_{\mu}\) and \(\omega_\mu\)
beyond the residual Lorentz freedom.  Having verified the static
representative at the level of Lorentz-scalar on-shell relations, we now
characterize the horizons selected by the same Casimir normalization.

\subsection{Horizon structure and Casimir mass function}
\label{subsec:horizon-structure-beta}

A Killing horizon is located at
\begin{equation}
\xi(r_h)=0.
\label{BH-horizon-general}
\end{equation}
Since \(e^{-2\beta r_h}\neq0\), Eq.~\eqref{BH-xi-beta} implies
\begin{equation}
\boxed{M_{\rm Cas}=-w_\beta(r_h).}
\label{BH-M-wh}
\end{equation}
Equivalently,
\begin{equation}
\boxed{
M_{\rm Cas}(r_h)=-\int_0^{r_h}ds\,e^{2\beta s}(\alpha s^2+\Lambda).
}
\label{BH-M-integral-beta}
\end{equation}
For \(\beta\neq0\), the integral may be written explicitly as
\begin{equation}
w_\beta(r)=
\alpha\left[
e^{2\beta r}\left(
\frac{r^2}{2\beta}-\frac{r}{2\beta^2}+\frac{1}{4\beta^3}
\right)-\frac{1}{4\beta^3}
\right]
+\frac{\Lambda}{2\beta}\left(e^{2\beta r}-1\right).
\label{BH-w-explicit-beta}
\end{equation}
The derivative of the Casimir-normalized Hamiltonian parameter function is
\begin{equation}
\frac{dM_{\rm Cas}}{dr_h}=-e^{2\beta r_h}(\alpha r_h^2+\Lambda)
=-e^{2\beta r_h}V(r_h).
\label{BH-dM-beta}
\end{equation}

The admissible parameter range follows directly from
\(m_\beta(r):=-w_\beta(r)\).  The restrictions introduced below are not
fixed by the local canonical analysis; they select a representative
non-extremal two-horizon branch of the Casimir mass function.  For the
static Killing-horizon branch
\begin{equation}
\alpha>0,
\qquad
\Lambda<0,
\label{BH-physical-branch}
\end{equation}
there is an extremal point at
\begin{equation}
r_*=\sqrt{-\Lambda/\alpha},
\qquad
V(r_*)=0,
\label{BH-extremal-radius}
\end{equation}
and the extremal Casimir mass is
\begin{equation}
M_{\rm ext}=m_\beta(r_*)=-w_\beta(r_*).
\label{BH-Mext-beta}
\end{equation}
For \(0<M_{\rm Cas}<M_{\rm ext}\), the equation
\(m_\beta(r_h)=M_{\rm Cas}\) has two roots,
\begin{equation}
r_-<r_*<r_+.
\label{BH-two-horizons}
\end{equation}
The metric is static where \(\xi(r)>0\), which is the interval
\((r_-,r_+)\) in this branch.  Within this static patch, \(r_-\) is the
inner Killing horizon and \(r_+\) has the character of an outer,
cosmological-type horizon.  At \(M_{\rm Cas}=M_{\rm ext}\) both roots
merge and the temperature vanishes.

For a representative point in parameter space, for example
\(\alpha=1\), \(\Lambda=-1\), \(\beta=0.2\), and
\(M_{\rm Cas}=0.4\), the condition \(0<M_{\rm Cas}<M_{\rm ext}\) is
satisfied and the horizon equation gives two roots \(r_-<r_*<r_+\).
This example is illustrative only.  The comparison with the
KV/Poisson--Sigma literature is analytic: it rests on the Casimir, the
mass function, and the extremal/non-extremal branch structure.

On a connected solution the Casimir takes the constant on-shell value
\begin{equation}
\mathcal C_{\rm DB}\big|_{\rm on\text{-}shell}=\mathcal C_0=-2M_{\rm Cas},
\label{BH-main-identity-beta}
\end{equation}
and at a horizon \(\xi(r_h)=0\) this reduces to
\(\mathcal C_0=2w_\beta(r_h)\), i.e.\ the horizon relation
\(M_{\rm Cas}=-w_\beta(r_h)\).  Thus \(M_{\rm Cas}\) is the
Casimir-normalized Hamiltonian parameter used in the static branch.  This
horizon relation fixes the variation of the Casimir-normalized mass
parameter with respect to the horizon position, and it is the input needed
for the surface-gravity and first-law normalization.

\subsection{Temperature and Casimir-normalized first law}
\label{subsec:thermodynamics-beta}

The surface gravity associated with the Killing time \(t\) must be
computed from the metric Killing norm \(N=e^{4\beta r}\xi\) in
\eqref{BH-Killing-norm}, rather than from the Casimir-normalized radial field
\(\xi\).  For a static frame \(ds^2=N\,dt^2-B\,dr^2\) with \(N(r_h)=0\), one has
\[
\kappa=\frac{|N'(r_h)|}{2\sqrt{NB}\,|_{r_h}}.
\]
Since \(NB=e^{4\beta r}\), \(\sqrt{NB}\,|_{r_h}=e^{2\beta r_h}\).  Moreover,
\(\xi(r_h)=0\) gives \(N'(r_h)=e^{4\beta r_h}\xi'(r_h)\), and
\(\xi'(r_h)=-2V(r_h)\) follows from~\eqref{BH-xi-prime-constraint-form}.
Therefore
\begin{equation}
\kappa_{\rm KV}=\frac{|N'(r_h)|}{2\sqrt{NB}\,|_{r_h}}
=\frac{e^{4\beta r_h}\,|\xi'(r_h)|}{2\,e^{2\beta r_h}}
=e^{2\beta r_h}\,|\alpha r_h^2+\Lambda|.
\label{BH-kappa-KV}
\end{equation}
Thus
\begin{equation}
\boxed{
T_{\rm KV}
=\frac{\kappa_{\rm KV}}{2\pi}
=\frac{e^{2\beta r_h}\,|\alpha r_h^2+\Lambda|}{2\pi}.
}
\label{BH-temperature-KV}
\end{equation}
In this normalization, torsion modifies the temperature through the factor
generated by \(\sqrt{NB}\neq1\).  Identifying the Killing norm with the
Casimir-normalized radial field \(\xi\) would correspond to a different
auxiliary normalization and would remove this torsional factor from the
temperature.
The torsionless value \(|\alpha r_h^2+\Lambda|/2\pi\) is recovered at
\(\beta=0\).

On the inner-horizon branch \(r_-=r_h<r_*\) one has \(V(r_h)<0\), and
Eq.~\eqref{BH-dM-beta} gives
\begin{equation}
dM_{\rm Cas}=e^{2\beta r_h}|V(r_h)|\,dr_h.
\label{BH-dM-KV-branch}
\end{equation}
Combining this with the temperature~\eqref{BH-temperature-KV}, the factor
\(e^{2\beta r_h}\) in \(dM_{\rm Cas}\) cancels against the same factor in
\(T_{\rm KV}\), so the first law \(dM_{\rm Cas}=T_{\rm KV}\,dS\) requires
\begin{equation}
dS=2\pi\,dr_h .
\label{BH-dS-KV}
\end{equation}
Choosing the additive constant so that \(S=0\) at \(r_h=0\),
\begin{equation}
\boxed{\ S=2\pi r_h=2\pi\varphi_h,\ }
\label{BH-entropy-KV}
\end{equation}
the standard two-dimensional dilaton entropy
\cite{wald1993black,iyer1994some,GegenbergKunstatterLouisMartinez1995}.  On this branch
\begin{equation}
\boxed{dM_{\rm Cas}=T_{\rm KV}\,dS.}
\label{BH-firstlaw-KV}
\end{equation}
In this normalization, torsion modifies the \emph{temperature}, but not
the horizon entropy.  For the outer horizon \(r_+\), the orientation of
the thermodynamic variation is reversed, as in de Sitter-type static
patches; the same formulae hold after using the corresponding horizon
orientation.

Because the metric Killing norm~\eqref{BH-Killing-norm} has been used,
\eqref{BH-entropy-KV} is consistent with the standard first-order
dilaton-gravity horizon entropy \(S=2\pi\varphi_h\).  In the Hamiltonian
normalization adopted here, the parameter entering the first-law relation
is the Casimir-normalized one, \(M_{\rm Cas}=-w_\beta(r_h)\); the
torsionless combination \(-\alpha r_h^3/3-\Lambda r_h\) coincides with it
only at \(\beta=0\), where it has the standard ADM interpretation.  The
field \(\xi\) is the Casimir-normalized metric field, not the Killing norm
of the on-shell geometry; using it as the Killing norm would not reproduce
the thermodynamic pair associated with the full static solution.\footnote{If one instead treated \(\xi\) as an auxiliary
Killing norm while keeping the same Casimir mass function
\(M_{\rm Cas}(r_h)=-w_\beta(r_h)\), the formal temperature would be
\(T_{\rm aux}=|V(r_h)|/(2\pi)\).  The corresponding first-law bookkeeping
entropy would then integrate to
\(S_{\rm aux}=\tfrac{\pi}{\beta}(e^{2\beta r_h}-1)\), up to an additive
constant.  This auxiliary entropy is not the first-order dilaton-gravity
horizon entropy of the on-shell metric \eqref{BH-Killing-norm}.}

With this normalization, the static-sector results chain together as
\begin{equation}
\underbrace{\partial_x\mathcal C_{\rm DB}\approx0}_{\text{constraint ideal}}
\quad\Longrightarrow\quad
\underbrace{M_{\rm Cas}:=-\tfrac12\mathcal C_{\rm DB}}_{\text{Casimir normalization}}
\quad\Longrightarrow\quad
\underbrace{dM_{\rm Cas}=T_{\rm KV}\,dS,\;\;S=2\pi\varphi_h}_{\text{first-law normalization}},
\label{BH-chain}
\end{equation}
with \(\beta\) entering the Casimir mass function, the Killing norm and
the KV-frame temperature, and the \(\beta\to0\) limit reproducing the
torsionless expressions.


\section{Conclusions and prospects}\label{sec:disc}

The extended Dirac--Bergmann analysis of the first-order KV model isolates
an auxiliary second-class sector whose elimination recovers the natural
canonical pairs of the action.  The resulting Dirac brackets coincide with
the generalized Faddeev--Jackiw brackets, establishing the equivalence of
the two reductions at the level of the reduced symplectic structure.

The reduced first-class constraint ideal also yields a gauge-unfixed
representative of the Katanaev/Poisson--Sigma Casimir.  This established
invariant is reconstructed within the canonical reduction, before any
conformal or light-cone gauge is imposed.  Its spatial derivative belongs
to the reduced constraint ideal, so the Casimir labels global solution
sectors without changing the local degree-of-freedom count.

The same canonical normalization was then applied to a static torsionful
sector.  In this branch, the Casimir-normalized field \(\xi\) fixes the
radial metric function through \(B=\xi^{-1}\), whereas the metric Killing
norm is \(N=e^{4\beta r}\xi\).  The geometry is therefore not in
Schwarzschild gauge for \(\beta\neq0\), and torsion modifies the Hawking
temperature through the normalization of the Killing time.  With the
standard first-order dilaton entropy, the inner-horizon branch satisfies
the Casimir-normalized first law \(dM_{\rm Cas}=T_{\rm KV}dS\).  The
canonical reconstruction thus fixes the static Hamiltonian normalization
and separates the Casimir-normalized radial field from the Killing norm
in the torsionful branch.

The construction is local on the spatial slice and does not replace a
boundary phase-space analysis.  Finite boundary charges, admissible
fall-off conditions and asymptotic symmetry algebras require additional
boundary data.  The present work supplies the reduced canonical
structure, the gauge-unfixed Casimir representative and the
static-sector normalization needed for such extensions, as well as for
applications to related KV-type models and to dilaton-matter couplings.

\backmatter

\bmhead{Acknowledgements}
The authors thank the Universidad Ju\'arez Aut\'onoma de Tabasco for
providing a suitable work environment while this research was carried
out.

\section*{Statements and Declarations}

\textbf{Funding.} J.M.C. acknowledges support from the Secretariat of
Science, Humanities, Technology, and Innovation (Secihti) of M\'exico
through a postdoctoral grant under Grant No.\ 3873825.

\textbf{Competing interests.} The authors declare no competing interests.

\textbf{Author contributions.} Both authors contributed to the
conceptualization, methodology, formal analysis, writing, review and
editing of the manuscript.  Both authors read and approved the final
manuscript.

\textbf{Data availability.} No datasets were generated or analysed during
the current study.

\textbf{Code availability.} No research code or numerical dataset was
generated or used during the current study.

\textbf{Use of AI-assisted tools.} AI-assisted tools were used for
language editing and consistency checks.  The authors reviewed and verified
all scientific claims, derivations, references and final wording, and take
full responsibility for the content of the manuscript.

\begin{appendices}
\renewcommand*{\theHequation}{Appendix.\Alph{section}.\arabic{equation}}

\section{Smeared rank check for the Dirac--Bergmann classification}
\label{app:rank-check}

This appendix verifies the constraint count used in
Sec.~\ref{sec:first-second-class}: the twelve-component constraint set
splits into six second-class and six first-class components, so that the
Poisson matrix \(W_E\) of Eq.~\eqref{24a} has rank and nullity equal to
six.  The computation is local on the spatial slice and uses only the
constraint algebra and the canonical brackets; no Euler--Lagrange
equation, gauge condition, or static-sector field equation is imposed.
To avoid manipulating bare derivatives of \(\delta(x-y)\), all
constraints are smeared with test functions and every delta-derivative is
moved onto the corresponding smearing functions by integration by parts.

The nine primary constraints of Eq.~\eqref{18}, together with the
secondary scalar \(\Psi\) and the secondary internal vector \(\Psi_I\),
are collected in
\begin{equation}
\mathcal Z[\mathbf f]=\int_\Sigma dx\,
\bigl(f\Phi+f^I\Phi_I+f^t_I\Phi^t_I+f^x_I\Phi^x_I
+f^t\Phi^t+f^x\Phi^x+h\Psi+h^I\Psi_I\bigr),
\label{app-smeared-Z}
\end{equation}
where
\[
\mathbf f=(f,f^I,f^t_I,f^x_I,f^t,f^x,h,h^I)
\]
is an arbitrary test multiplet adapted to the ordering
\[
(\Phi,\Phi_I,\Phi^t_I,\Phi^x_I,\Phi^t,\Phi^x,\Psi,\Psi_I).
\]
Equivalently, \(W_E\) is represented by the bilinear form
\(\{\mathcal Z[\mathbf f],\mathcal Z[\mathbf g]\}\), after all derivatives of Dirac delta
distributions have been transferred to the smearing functions.

\emph{Second-class block (rank six).}
We first test the auxiliary six-component set
\[
\zeta=(\chi,\chi_I,\chi^x,\chi^x_I),
\]
defined in Eq.~\eqref{27}, as a candidate second-class sector.  The
criterion is algebraic: the Poisson matrix restricted to this subspace,
\[
C^{\alpha\beta}(x,y)=\{\zeta^\alpha(x),\zeta^\beta(y)\},
\]
must be non-degenerate on the constraint surface.  If this restricted
matrix has maximal rank in the six-dimensional auxiliary subspace, then
\(\zeta\) provides a non-degenerate complement to the kernel of the full
matrix \(W_E\).

The set \(\zeta\) splits into two canonically conjugate triplets,
\[
(\chi,\chi_I)
\qquad\text{and}\qquad
(\chi^x,\chi^x_I),
\]
where the first triplet contains the auxiliary momenta and the second
contains the corresponding coordinate constraints.  Using
\(\chi=\Pi\), \(\chi^x=\Pi^x-\varphi\),
\(\chi_I=\Pi_I\), \(\chi^x_I=\Pi^x_I-\varphi_I\), together with the
canonical brackets
\[
\{\varphi(x),\Pi(y)\}=\delta(x-y),
\qquad
\{\varphi_I(x),\Pi_J(y)\}=\eta_{IJ}\delta(x-y),
\]
the only non-vanishing pairings between the two triplets are
\begin{align}
\{\chi[f],\chi^x[g]\}
&=
-\int_\Sigma dx\,dy\,f(x)g(y)\,\{\Pi(x),\varphi(y)\}
=
\int_\Sigma dx\,f g,
\nonumber\\
\{\chi_I[f^I],\chi^x_J[g^J]\}
&=
-\int_\Sigma dx\,dy\,f^I(x)g^J(y)\,\{\Pi_I(x),\varphi_J(y)\}
=
\int_\Sigma dx\,f^I g_I.
\label{app-aux-block}
\end{align}
All brackets internal to a single triplet vanish.  This is the smeared
form of the non-degenerate block \(C^{\alpha\beta}\) displayed in
Eq.~\eqref{29}.  In the paired basis
\((\chi,\chi_I;\chi^x,\chi^x_I)\), the auxiliary block has the local
form
\[
\Omega_{\rm aux}
=
\begin{pmatrix}
0 & \mathsf G\\
-\mathsf G & 0
\end{pmatrix},
\qquad
\mathsf G=\operatorname{diag}(1,\eta_{IJ}).
\]
Since \(\mathsf G\) is non-degenerate, \(\Omega_{\rm aux}\) is
non-degenerate.  Hence the restricted matrix has maximal rank,
\[
\operatorname{rank}C^{\alpha\beta}
=
\operatorname{rank}\Omega_{\rm aux}
=
6.
\]
Consequently, the auxiliary constraints are genuinely second class: no
non-trivial linear combination within the subspace spanned by
\(\zeta\) lies in the kernel of \(W_E\).  This gives
\[
\operatorname{rank}W_E\geq6.
\]

\emph{First-class kernel (nullity six).}
The remaining directions are the null directions of \(W_E\).  In the main
text they are displayed in Eq.~\eqref{null-vectors-display} as the four
tensorial families
\[
V^{(1)},\qquad
V^{(2)}{}_{I},\qquad
V^{(3)}{}_{I},\qquad
V^{(4)}.
\]
Their component count is determined by their free internal indices:
\(V^{(1)}\) and \(V^{(4)}\) carry no free internal index and therefore
contribute one direction each, whereas \(V^{(2)}{}_{I}\) and
\(V^{(3)}{}_{I}\) carry one free internal index \(I=0,1\) and therefore
contribute two directions each.  Thus
\[
1+2+2+1=6.
\]
Contracting these null vectors with the complete constraint multiplet
reproduces the first-class combinations of Eq.~\eqref{26},
\begin{equation}
\gamma_A=(\gamma,\gamma_I,\gamma^t_I,\gamma^t),
\qquad
1+2+2+1=6.
\label{app-null-combinations}
\end{equation}
Here \(\gamma\) is the Lorentz-scalar combination, \(\gamma_I\) is the
internal Lorentz-vector family, \(\gamma^t_I\) is the temporal
internal-vector family, and \(\gamma^t\) is the temporal scalar.  The
temporal generators are primary first-class constraints:
\[
\gamma^t=\Phi^t,
\qquad
\gamma^t_I=\Phi^t_I,
\]
namely the momenta conjugate to the multipliers \(\omega_t\) and
\(e^I_t\).  For \(\gamma\) and \(\gamma_I\), the auxiliary terms in
Eq.~\eqref{26} cancel the brackets of the secondary constraints with the
auxiliary sector \((\chi,\chi_I,\chi^x,\chi^x_I)\).  After the
delta-derivatives are integrated by parts, the remaining terms are
proportional to constraints.  Therefore, for an arbitrary test multiplet
\(\mathbf u\),
\begin{equation}
\{\gamma_A[\epsilon^A],\mathcal Z[\mathbf u]\}\approx0.
\label{app-null-test}
\end{equation}
Thus the six independent components of \(\gamma_A\) lie in the kernel of
\(W_E\).  Hence
\[
\operatorname{nullity}W_E\geq6.
\]

The argument now closes by the rank-nullity relation for the
twelve-component constraint matrix.  The non-degenerate auxiliary block
gives \(\operatorname{rank}W_E\geq6\).  The six independent combinations
\(\gamma_A\) lie in the kernel of \(W_E\), so
\(\operatorname{nullity}W_E\geq6\), or equivalently
\(\operatorname{rank}W_E\leq6\).  Therefore
\begin{equation}
\operatorname{rank}W_E=6,\qquad
\operatorname{nullity}W_E=6.
\label{app-rank-nullity}
\end{equation}
The auxiliary set \(\zeta\) is a non-degenerate second-class complement
to the first-class kernel spanned by \(\gamma_A\).  This is the \(6+6\)
classification used in Table~\ref{tab:constraint-classification}: six
second-class components in \(\zeta\) and six first-class components in
\(\gamma_A\).  The check rests only on the constraint algebra and the
canonical brackets, with no dynamical input beyond the Dirac--Bergmann
consistency analysis developed in the main text.

\section{Details of the Faddeev--Jackiw symplectic matrices}\label{app:FJ-matrices}

This appendix records the technical steps used in
Sec.~\ref{sec:FJ}.  The matrices are written in block notation, with
internal indices carried by the corresponding entries.  They are not
needed for the main line of the argument, but they make explicit the
zero-mode test and the inversion leading to the generalized FJ brackets.

After the first zero-mode contraction gives the constraints
$\Omega^{(0)}$ and $\Omega^{(0)}_I$, the enlarged rectangular matrix used
to test for further constraints is
\begin{equation}
\begingroup
\setlength{\arraycolsep}{2pt}
\resizebox{0.98\textwidth}{!}{$\displaystyle
f^{(1)}_{cb}=
\left(
 \begin{array}{ccc}
   f^{(0)}_{ab}\\
   \frac{\partial\Omega^{(0)}}{\partial\xi^{(0)b}}\\
   \frac{\partial\Omega^{(0)}_{I}}{\partial\xi^{(0)b}}
  \end{array}
\right)=\left(
 \begin{array}{cccccc}
   0&0&0&1&0&0\\
   0&0&0&0&0&\delta^{I}_{J}\\
   0&0&0&0&0&0\\
   -1&0&0&0&0&0\\
   0&0&0&0&0&0\\
   0&-\delta^{J}_{I}&0&0&0&0\\
   \partial_{x}&\varepsilon{^{J}}_{K}e^{K}_{x}&0&0&0&\varepsilon{^{I}}_{J}\varphi_{J}\\
   -2\alpha\varphi\varepsilon_{IK}e^{K}_{x}&\delta^{J}_{I}\partial_{x}+\varepsilon{_{I}}^{J}\omega_{x}-2\beta\varepsilon_{IK}e^{K}_{x}\varphi^{J}&0&\varepsilon{_{I}}^{J}\varphi_{J}&0&-\varepsilon_{IJ}E
  \end{array}
\right)\delta(x-y).
$}
\endgroup
\label{app-FJ-rectangular}
\end{equation}
A convenient set of zero modes is
\begin{align}
(v^{(1)})_{1}^{T}
&=\bigl(0,\varepsilon{_{I}}^{J}\varphi_{J}\delta(x-y),0,
\partial_{x}\delta(x-y),
0,\varepsilon{^{I}}_{K}e^{K}_{x}\delta(x-y),
\delta(x-y),0\bigr),
\nonumber\\
(v^{(1)})_{2}^{T}
&=\bigl(\varepsilon{_{I}}^{K}\varphi_{K}\delta(x-y),
\varepsilon_{IJ}E\delta(x-y),0,
-2\alpha\varepsilon_{IK}e^{K}_{x}\delta(x-y),
\nonumber\\
&\quad 0,\delta^{J}_{I}\partial_{x}\delta(x-y)
+\varepsilon{_{I}}^{J}\omega_{x}\delta(x-y)
-2\beta\varepsilon_{IK}e^{K}_{x}\varphi^{J}\delta(x-y),
0,\delta^{J}_{I}\delta(x-y)\bigr).
\label{app-FJ-zero-modes}
\end{align}
Contracting these zero modes with the vector $Z_c$ of the FJ equations,
\begin{equation}
(v^{(1)})^{T}_{c}Z_c=0,
\end{equation}
one obtains an identity.  Therefore the FJ algorithm produces no further
constraints beyond Eq.~\eqref{45}.

After incorporating the constraints through the multipliers
$\lambda$ and $\lambda^I$, the symplectic variables are
\begin{equation}
\xi^{(1)a}=\{\varphi,\varphi_I,\lambda,\omega_x,
\lambda^I,e^I_x\},
\end{equation}
with one-forms
\begin{equation}
a^{(1)}_a=\{0,0,\Omega^{(0)},\varphi,
\Omega^{(0)}_I,\varphi_I\}.
\end{equation}
The resulting presymplectic matrix is
\begin{equation}
\begingroup
\text{\scriptsize}
\setlength{\arraycolsep}{2pt}
\renewcommand{\arraystretch}{1.05}
\begin{adjustbox}{max width=0.98\textwidth,center}
$\displaystyle
f^{(1)}_{ab}(x,y) =
\left(
\begin{array}{cccccc}
0 & 0 & \partial_{y} & 1 & M_{15} & 0 \\
0 & 0 & \varepsilon^{I}_{\ K}e^{K}_{y} & 0 & M_{25} & \delta^{I}_{J} \\
-\partial_{x} & -\varepsilon^{J}_{\ K}e^{K}_{x} & 0 & 0 & 0 & -\varepsilon^{K}_{\ J}\varphi_{K} \\
-1 & 0 & 0 & 0 & \varepsilon_{J}^{\ L}\varphi_{L} & 0 \\
M_{51} & M_{52} & 0 & -\varepsilon_{I}^{\ K}\varphi_{K} & 0 & \varepsilon_{IJ}E \\
0 & -\delta^{J}_{I} & \varepsilon^{K}_{\ I}\varphi_{K} & 0 & \varepsilon_{IJ}E & 0
\end{array}
\right)\delta(x-y),
$
\end{adjustbox}
\endgroup
\label{app-FJ-f1}
\end{equation}
where
\begin{align}
M_{15} &= -2\alpha\varepsilon_{JK}e^{K}_{y}\varphi,
&
M_{25} &= \delta^{I}_{J}\partial_{y} - \varepsilon^{I}_{\ J}\omega_{y}
 - 2\beta\varepsilon_{JK}e^{K}_{y}\varphi^{I},
\nonumber\\
M_{51} &= 2\alpha\varphi\varepsilon_{IK}e^{K}_{x},
&
M_{52} &= -\delta^{J}_{I}\partial_{x} - \varepsilon_{I}^{\ J}\omega_{x}
 + 2\beta\varepsilon_{IK}e^{K}_{x}\varphi^{J}.
\end{align}
This matrix is singular because it still contains gauge directions.  The
temporal gauge conditions for the original multiplier sector are then
implemented by adding the multiplier variables $\theta$ and $\alpha_I$,
leading to
\begin{equation}
\mathcal{L}^{(2)}=
\varphi\dot{\omega}_{x}+
\varphi_{I}\dot{e}^{I}_{x}+
\left(\Omega^{(0)}_{I}+\alpha_{I}\right)\dot{\lambda}^{I}+
\left(\Omega^{(0)}+\theta\right)\dot{\lambda}.
\label{app-FJ-L2}
\end{equation}
For
\begin{equation}
\xi^{(2)a}=\{\varphi,\varphi_I,\lambda,\theta,
\omega_x,\lambda^I,\alpha_I,e^I_x\},
\end{equation}
the final symplectic matrix is
\begin{equation}
\begingroup
\text{\scriptsize}
\setlength{\arraycolsep}{2pt}
\renewcommand{\arraystretch}{1.05}
\begin{adjustbox}{max width=0.98\textwidth,center}
$\displaystyle
f^{(2)}_{ab}(x,y) =
\left(
\begin{array}{cccccccc}
0 & 0 & \partial_{y} & 0 & 1 & M_{16} & 0 & 0 \\
0 & 0 & \varepsilon^{I}_{\ K}e^{K}_{y} & 0 & 0 & M_{26} & 0 & \delta^{I}_{J} \\
-\partial_{x} & -\varepsilon^{J}_{\ K}e^{K}_{x} & 0 & -1 & 0 & 0 & 0 & N_{38} \\
0 & 0 & 1 & 0 & 0 & 0 & 0 & 0 \\
-1 & 0 & 0 & 0 & 0 & \varepsilon_{J}^{\ L}\varphi_{L} & 0 & 0 \\
M_{61} & M_{62} & 0 & 0 & N_{65} & 0 & -\delta^{J}_{I} & \varepsilon_{IJ}E \\
0 & 0 & 0 & 0 & 0 & \delta^{I}_{J} & 0 & 0 \\
0 & -\delta^{J}_{I} & \varepsilon^{K}_{\ I}\varphi_{K} & 0 & 0 & \varepsilon_{IJ}E & 0 & 0
\end{array}
\right)\delta(x-y),
$
\end{adjustbox}
\endgroup
\label{app-FJ-f2}
\end{equation}
with
\begin{align}
M_{16} &= -2\alpha\varepsilon_{JK}e^{K}_{y}\varphi,
&
M_{26} &= \delta^{I}_{J}\partial_{y} - \varepsilon^{I}_{\ J}\omega_{y}
 - 2\beta\varepsilon_{JK}e^{K}_{y}\varphi^{I},
\nonumber\\
M_{61} &= 2\alpha\varphi\varepsilon_{IK}e^{K}_{x},
&
M_{62} &= -\delta^{J}_{I}\partial_{x} - \varepsilon_{I}^{\ J}\omega_{x}
 + 2\beta\varepsilon_{IK}e^{K}_{x}\varphi^{J},
\nonumber\\
N_{38} &= -\varepsilon^{K}_{\ J}\varphi_{K},
&
N_{65} &= -\varepsilon_{I}^{\ K}\varphi_{K}.
\end{align}
The inverse of $f^{(2)}_{ab}$ contains, in particular, the entries
\begin{equation}
\begin{split}
\left[f^{(2)}\right]^{-1}_{\omega_x\varphi}(x,y)&=\delta(x-y),\\
\left[f^{(2)}\right]^{-1}_{e^J_x\varphi_I}(x,y)&=\delta^J_I\delta(x-y),
\end{split}
\label{app-FJ-inverse-relevant}
\end{equation}
which are the entries used in Eq.~\eqref{56}.  A representative block
form of the inverse, restricted to the same ordering of variables, is
\begin{equation}
\begingroup
\setlength{\arraycolsep}{2pt}
\resizebox{0.98\textwidth}{!}{$\displaystyle
[f^{(2)}_{ab}(x,y)]^{-1}=
\left(
\begin{array}{cccccccc}
0 & 0 & 0 & 0 & -1 & 0 & \varepsilon_{I}^{\ K}\varphi_{K} & 0 \\
0 & 0 & 0 & \varepsilon^{K}_{\ J}\varphi_{K} & 0 & 0 & \varepsilon_{JI}E & -\delta^{I}_{J} \\
0 & 0 & 0 & 1 & 0 & 0 & 0 & 0 \\
0 & -\varepsilon^{K}_{\ J}\varphi_{K} & -1 & 0 & -\partial_{y} & 0 & 0 & \varepsilon^{I}_{\ K}e^{K}_{y} \\
1 & 0 & 0 & \partial_{x} & 0 & 0 & 2\alpha\varphi\varepsilon_{JK}e^{K}_{x} & 0 \\
0 & 0 & 0 & 0 & 0 & 0 & \delta^{J}_{I} & 0 \\
-\varepsilon_{J}^{\ K}\varphi_{K} & \varepsilon_{JI}E & 0 & 0 & -2\alpha\varphi\varepsilon_{JK}e^{K}_{x} & -\delta^{I}_{J} & 0 & G \\
0 & \delta^{J}_{I} & 0 & -\varepsilon^{J}_{\ K}e^{K}_{y} & 0 & 0 & H & 0
\end{array}
\right)\delta(x-y),
$}
\endgroup
\label{app-FJ-inverse}
\end{equation}
where
\begin{align}
G &= \delta^{I}_{J}\partial_{x} - \varepsilon_{J}^{\ I}\omega_{x}
 - 2\beta\varepsilon_{JL}e^{L}_{x}\varphi^{I},
\nonumber\\
H &= -\delta^{J}_{I}\partial_{y} - \varepsilon_{I}^{\ J}\omega_{y}
 + 2\beta\varepsilon_{IK}e^{K}_{y}\varphi^{J}.
\end{align}
The displayed inverse is used only to extract the reduced brackets and
to verify their agreement with the Dirac brackets.  The gauge fixing
introduced to make $f^{(2)}_{ab}$ invertible is not part of the
gauge-unfixed derivation of the Casimir in Sec.~\ref{sec:casimir}.


\section{On-shell residuals of the torsionful static representative}
\label{app:torsionful-residuals}

This appendix records the component identities used in
Sec.~\ref{subsec:torsionful-onshell-checks}.  In the static patch
\(\xi>0\) we use
\begin{equation}
V(r)=\alpha r^2+\Lambda,
\qquad
E(r)=V(r)+\beta\xi(r),
\qquad
\xi'(r)=-2E(r),
\label{app-static-identities}
\end{equation}
and the representative
\begin{equation}
N=e^{4\beta r}\xi,
\qquad
B=\xi^{-1},
\qquad
e=\sqrt{NB}=e^{2\beta r},
\label{app-static-metric}
\end{equation}
with
\begin{equation}
e^0=e^{2\beta r}\sqrt{\xi}\,dt,
\qquad
e^1=\xi^{-1/2}\,dr,
\qquad
\varphi=r,
\qquad
\varphi_0=-\sqrt{\xi},
\qquad
\varphi_1=0,
\label{app-static-fields}
\end{equation}
and
\begin{equation}
\omega=\omega_t(r)\,dt,
\qquad
\omega_t=-e^{2\beta r}E,
\qquad
\omega_r=0.
\label{app-static-connection}
\end{equation}
The sign of \(\varphi_0\) fixes the internal orientation of this local
representative; the Lorentz scalars used in the main text are insensitive
to this choice.

Among the four field equations---the curvature (\(\varphi\)-), connection
(\(\omega\)-), torsion (\(\varphi_I\)-) and zweibein (\(e^I\)-)
equations---the curvature equation provides a particularly transparent
check, since it closes only once the non-Schwarzschild relation
\(NB=e^{4\beta r}\) is used.  With the orientation used in the main text,
\(F=-\partial_r\omega_t\).  Hence
\begin{align}
F-2\alpha e r
&=\partial_r(e^{2\beta r}E)-2\alpha e^{2\beta r}r
\nonumber\\
&=e^{2\beta r}\left(2\beta E+E'-2\alpha r\right)
\nonumber\\
&=e^{2\beta r}\left(2\beta E+2\alpha r+\beta\xi'-2\alpha r\right)
\nonumber\\
&=e^{2\beta r}\beta(2E+\xi')=0.
\label{app-curvature-residual}
\end{align}
Thus the \(\varphi\)-equation is satisfied identically once
\eqref{app-static-identities} is used.

The connection equation
\begin{equation}
\partial_\mu\varphi+\varepsilon^{IJ}\varphi_Ie_{\mu J}=0
\label{app-omega-residual}
\end{equation}
reduces to \(\varphi_1=0\) for \(\mu=t\) and to
\(\varphi_0=-\sqrt{\xi}\) (equivalently \(\varphi_0^2=\xi\)) for
\(\mu=r\), in agreement with the assignments in
\eqref{app-static-fields}.  Only the scalar \(\varphi_0^2=\xi\) enters the
Lorentz invariants of the main text, so the sign is a matter of internal
orientation.  The \(\varphi_I\)-equations give
\begin{equation}
T^I-2\beta e\varphi^I=0,
\label{app-torsion-residual}
\end{equation}
and direct substitution of \eqref{app-static-metric}--\eqref{app-static-connection}
reduces the two independent components to identities.  Equivalently,
for the normalized torsion \(\mathcal T^I=e^{-1}T^I\), one obtains
\begin{equation}
\mathcal T_I\mathcal T^I=4\beta^2\varphi_I\varphi^I=4\beta^2\xi.
\label{app-normalized-torsion}
\end{equation}
Finally, the independent components of the zweibein equations are, up to
non-zero frame factors, proportional to
\begin{equation}
\xi'+2E,
\qquad
\omega_t+e^{2\beta r}E,
\qquad
\varphi_1,
\label{app-frame-residuals}
\end{equation}
which vanish by \eqref{app-static-identities}, \eqref{app-static-fields}
and \eqref{app-static-connection}.  These residuals show explicitly why
the non-Schwarzschild factor \(NB=e^{4\beta r}\) is required in the
torsionful sector.

\end{appendices}

\bibliography{KV_bibliography}

\end{document}